\documentclass[11pt,a4paper]{article}

\usepackage{a4wide}
\usepackage[english]{babel}   
\usepackage{epsfig}
\usepackage{amsfonts}
\usepackage{bbm}
\usepackage{graphicx}
\usepackage{subfig}
\usepackage{amsmath}
\usepackage{float}

\usepackage{listings}
\newcommand{\be}{\begin{eqnarray}}
\newcommand{\ee}{\end{eqnarray}}
\newcommand{\beqn}{\begin{eqnarray}}
\newcommand{\eeqn}{\end{eqnarray}}
\newcommand{\bes}{\begin{eqnarray*}}
\newcommand{\ees}{\end{eqnarray*}}
\newcommand{\beqns}{\begin{eqnarray*}}
\newcommand{\eeqns}{\end{eqnarray*}}

\newcommand{\rmd}{\mbox{d}}

\newcommand{\dd}[2]{\frac{\rmd{#1}}{\rmd{#2}}}

\newcommand{\bfn}[1]{\mbox{\protect\bf #1}}

\begin{document}

\title{Rigidly rotating gravitationally bound systems\\
  of point particles, compared to polytropes}
\author{Yngve Hopstad and Jan Myrheim,\\
Department of Physics, NTNU, N--7491 Trondheim, Norway}

\maketitle

\begin{abstract}
  In order to simulate rigidly rotating polytropes we have simulated
  systems of $N$ point particles, with $N$ up to 1800.  Two particles
  at a distance $r$ interact by an attractive potential $-1/r$ and a
  repulsive potential $1/r^2$.  The repulsion simulates the pressure
  in a polytropic gas of polytropic index $3/2$.  We take the total
  angular momentum $L$ to be conserved, but not the total energy $E$.
  The particles are stationary in the rotating coordinate system.  The
  rotational energy is $L^2/(2I)$ where $I$ is the moment of inertia.
  Configurations where the energy $E$ has a local minimum are stable.
  In the continuum limit $N\to\infty$ the particles become more and
  more tightly packed in a finite volume, with the interparticle
  distances decreasing as $N^{-1/3}$.  We argue that $N^{-1/3}$ is a
  good parameter for describing the continuum limit.  We argue further
  that the continuum limit is the polytropic gas of index $3/2$.  For
  example, the density profile of the nonrotating gas approaches that
  computed from the Lane--Emden equation describing the nonrotating
  polytropic gas.  In the case of maximum rotation the instability
  occurs by the loss of particles from the equator, which becomes a
  sharp edge, as predicted by Jeans in his study of rotating
  polytropes.  We describe the minimum energy nonrotating
  configurations for a number of small values of $N$.

  $keywords$: polytropes; rotation; numerical simulation; virial
  theorem; Lane--Emden equation; Jeans effect.
\end{abstract}

\section{Introduction}
\label{sec:1}

In a previous article, referred to here as paper I, we studied
rotating systems of up to five particles held together by a long range
attractive potential, inversely proportional to the distance, and
stabilized by a short range repulsive potential of inverse square
type~\cite{HM}.  As explained there, our motivation was to use the
same model, with many particles, to simulate nonrotating and rigidly
rotating polytropic gases, described by an equation of state of the
form
\be
\label{eq:polystate}
P=K\rho^{\gamma}\;,\qquad
\gamma=1+\frac{1}{n}\;,
\ee
where $P$ is the pressure and $\rho$ the mass density.  Here $K$,
$\gamma$, and $n$ are constants, $\gamma$ is called the adiabatic
index and $n$ is the polytropic index.  The nonrotating polytropes are
described by the Lane--Emden equation, which can also be generalized
to the rotating case. 

We report here results of simulations with up to 1800 particles, and
argue that this discrete model does indeed behave much like a
continuous polytrope with polytropic index $n=3/2$.  Our methods for
generating random equilibrium configurations and calculating their
stability are described in paper I, therefore we summarize more
briefly here.

It is possible to simulate different values of the adiabatic index by
changing the power of the repulsive potential.  Note however that a
power different from the inverse square would introduce a minor
complication in treating the virial theorem.  In the present article
we have studied only the inverse square potential.

The early history of the theory of nonrotating gravitationally bound
gaseous bodies is nicely summarized by Chandrasekhar, in the form of
bibliographical notes to Chapter~IV in his book on stellar
structure~\cite{ChandraStellarStr}.  Pioneers were especially
J.H.~Lane, Lord Kelvin, A.~Ritter, and R.~Emden.

It seems that Emden was among the first to express the Lane--Emden
equation in the dimensionless form we know today, in his
{\em Gaskugeln} from 1907~\cite{Emd}.  In this work he explored the
various solutions it gives for different values of $n$, and used these
solutions to calculate the central values of pressure, density and
temperature of the Sun and some stars, as well as the structure of the
atmosphere of the Earth.  He also discussed the rotation of the Sun
and its pulsation.

In 1916 Eddington presented what is now called the Eddington Standard
Model of stars~\cite{Edd}.  He included the radiation pressure in his
equation of state.  In a massive star, where this dominates, it gives
$\gamma=4/3$ and $n=3$.  He could then use the results obtained by
Emden from the Lane--Emden equation with $n=3$.

Eddington did not include rotation in his equations, but in 1923
E.A.~Milne included slow rotation in an extended Lane--Emden equation
for $n=3$, which is then Eq.~(\ref{eq:ChaMilne}) with
$\omega\neq 0$~\cite{Mil}.  Milne argued that the luminosity of a
rotating star is not the same along the different principal axes.
This argument led H.~von Zeipel to a well known theorem on the
influence of rotation on the energy output of a star, known today as
gravity darkening~\cite{Esp}.  The extended equation of Milne gave
Chandrasekhar the idea to develop a series expansion for solutions of
the Lane--Emden equation for slowly rotating polytropes with the
angular velocity as expansion parameter~\cite{Cha}.  This expansion
has ever since served as the starting point for other similar
expansions that try to describe slow and fast rotating polytropes and
stars.  Because Chandrasekhar's work is heavily built on Milne's work
it is called the Chandrasekhar--Milne expansion.

A different approach to the study of rotating polytropes is that of
J.H.~Jeans, for which he recieved the Adams Prize in 1917.  Some of
his results, that we call here the Jeans effect, are partly confirmed
by our work.  He concluded that there exists a critical value of the
index,
\be
\gamma_c=1+\frac{1}{n_c}\;,
\ee
such that for $\gamma<\gamma_c$, i.e.~$n>n_c$, the polytrope behaves
differently from an incompressible fluid, in the following ways.
 \begin{center}
  \begin{itemize}
  \item The Jacobi transition, breaking the rotational symmetry in the
    rotation plane as the angular momentum increases, does not happen.
  \item The rotating polytrope becomes unstable by losing particles
    from its equator when the centrifugal force there exceeds the
    gravitational attraction.
  \item At the critical angular momentum, where the instability sets in,
    the equator line becomes a sharp edge.
  \end{itemize}
\end{center}
In Appendix~\ref{app:Jeans} we have given a brief summary of how Jeans
derived these results.  He estimated that
\be
\gamma_c\approx 2.2\;,\qquad n_c\approx 0.83\;.
\ee
Our values of $\gamma=5/3$, $n=3/2$ are well within the regime of the
Jeans effect, $0.83<n<5$.

In the theory of rotating incompressible fluids ($\gamma\to\infty$)
the Jacobi transition is the change, with increasing angular momentum,
from a Maclaurin ellipsoid, having two long axes of equal length, to a
Jacobi ellipsoid, with three different axes.  Above the transition,
both shapes exist as equilibrium configurations, but the Jacobi
ellipsoid has lower energy, hence it should appear in our simulations
where we minimize the energy.  Jeans found that this transition is the
leading instability down to $\gamma=\gamma_c$.  For $\gamma<\gamma_c$
the leading instability is the loss of particles due to the
centrifugal force.

Jeans approached the problem of rotating polytropes in what he called
the adiabatic model.  He expanded the density as a series in a
compressibility factor $\epsilon$~\cite{JeansI,JeansII,JeansIII}.
This allowed him to express the effective potential (gravitational
plus centrifugal) as a series expanion, an idea which seems to have
come from Lyapunov in the first place.  See~\cite{Kho} for a brief
summary of Lyapunov's work on celestial bodies.  The methods of Jeans
and Lyapunov were different in several ways.  For example, Jeans used
Cartesian coordinates, while Lyapunov used spherical coordinates,
Lam{\'e} functions and Legendre polynomials, combined with his own
invented series and techniques.

The minimum value $\epsilon=0$ represents an incompressible fluid,
which has a surface density equal to its central density, and the
maximum value $\epsilon=1$ represents a polytrope, which has a surface
density equal to zero.  By thinking of $\epsilon$ as varying
continuously we may imagine a smooth transition between the two
extremes.  His expansion only to first order in $\epsilon$ gives
surprisingly accurate results, as confirmed by the work of James in
1961, who concluded that
\be
0.808<n_c<0.8085\;.
\ee
In order to verify his results for polytropes, Jeans introduced what
he called the generalized Roche model, which is an approximation where
most of the mass is contained in an incompressible inner region.  In
the Roche model, the inner region is just a point.

The method used by James was very different, starting with the
Chandrasekhar--Milne expansion and further expanding the terms in this
expansion in Legendre polynomials~\cite{James}.  Another work on fast
rotating polytropes using the Chandrasekhar--Milne expansion is that
of Monaghan and Roxburgh~\cite{Rox}. They allow the inner region to be
compressible, unlike in the generalized Roche model of Jeans.

A different approach, based on a variational principle, is that of
Hurley and Roberts~\cite{RobIandII,RobIII}.  A recent work, where
these results are verified numerically, is that of Kong et
al.~\cite{Kongetal}.

\subsection{Remarks on discrete models for continuous systems}

Our model for the gravitationally bound rotating polytropic gas is a
system of $N$ point particles interacting by a long range attractive
and a short range repulsive potential, static in the rotating
reference system.  Given a fixed value of the angular momentum, we
look for local minima of the potential plus rotational energy.  We
choose this approach because it is a much simpler problem,
computationally, than solving the partial differential equations
describing a fluid.  It is useful to know that the approach to the
continuum limit is rather slow, we find that it is asymptotically
linear in $N^{-1/3}$.

It is a complication that there exist a very large number of local
minima, separated by low barriers of energy.  For this reason we use
Monte Carlo methods when searching for minima.  Fortunately, we need
not find all the minima, because the different minima are presumably
equivalent descriptions of the same minimum energy configuration of
the continuous fluid.  The energy barriers make our model system
rigid, so that we can only simulate rigid rotation.

Thus we may simulate the rotation of white dwarfs, but not the
differential rotation of stars as treated by the ESTER
code~\cite{MRFELBP}.  In order to simulate neutron stars we would have
to take into account general relativistic effects, and it is not
immediately clear how to do so~\cite{DGREG}.

Obviously, our method may be useful for simulating rigidly rotating
systems of finite numbers of particles.  Asteroids that are
gravitationally bound assemblies of rocks may be simulated if we
introduce hard core repulsive potentials.  Simulations of quantum
systems, such as large molecules and atomic nuclei, would give both
the moments of inertia and the vibrational frequencies needed for
computing rotational and vibrational spectra.

A different approach to the study of gravitationally bound systems is
taken for example by Chavanis and Rieutord, who consider fermions at
nonzero temperature as a model of elliptical galaxies and globular
clusters~\cite{CR}.  They have to enclose the system in a box to
prevent evaporation.  Postulating a Fermi--Dirac distribution is a way
of handling the kinetic energy in order to prevent collapse.  It works
also in the limit of zero temperature, where the fermion gas is a
polytrope of index $n=3/2$.

\subsection{Outline of the article}

In Sec.~\ref{sec:2} we present some general theory, including some
theory from paper I in order to make the present article
self-contained.

Before discussing the rotating configurations we take a closer look at
the nonrotating case.  In Sec.~\ref{sec:3} we study how some
properties of minimum energy configurations depend on $N$, the number
of particles.  We find that $N^{-1/3}$ is a very useful expansion
parameter, in particular because it goes to zero in the continuum
limit $N\to\infty$.  It appears that several quantities describing the
configurations take finite values in this limit.  One example is the
average energy per particle pair, which is fitted remarkably well all
the way down to $N=2$ by a cubic polynomial in $N^{-1/3}$.  The fit
indicates that $N=1800$, which is the largest system we have
simulated, is rather far from the continuum limit.  In fact, one
should remember that the convergence of $N^{-1/3}$ to zero is rather
slow.

Similar examples are two quantities defining the geometrical size of a
configuration.  The result that the size remains finite as
$N\to\infty$ can be understood as a consequence of the virial theorem.

The Lane--Emden equation describes the density profile of a
gravitationally bound nonrotating polytropic gas.  In our system of a
finite number of particles we can compute an approximate density
profile that compares very well with the one predicted from the
continuum theory.

In Sec.~\ref{sec:4} we present plots of minimum energy configurations
of few particles, up to $N=15$, $N=26$, and $N=53$.  For some special
values of $N$ the particles can be arranged in particularly symmetric
configurations, and it is no surprise that these are ``magic numbers''
for which the energy is particularly low,

Section~\ref{sec:5} is a brief general introdution to the case of
rotating systems.  We find a power law for the dependence of the
maximal angular momentum on the number of particles, with a power
which is close to, but apparently different from $1.5$.  We argue
that the power should go to $1.5$ as $N\to\infty$.

In Sec.~\ref{sec:6} we present a detailed study of rotating systems of
400 particles.  Some other examples with different numbers of
particles are included for comparison.  The results may be summarized
as follows.

The stable configurations with different values of the angular
momentum $L$ are arranged in a very large number of branches.  Each
branch is stable within a limited $L$ interval, and becomes unstable
at either end of the interval by changing discontinuously into a
different configuration.

We have followed many branches by varying the angular momentum in
small steps.  It is very clear, however, that our sample of stable
branches is very far from complete.  To demonstrate this we have
searched for and found other stable configurations at one randomly
chosen value of the angular momentum, $L=3067$.

We present plots of the stability parameter $\sigma$, defined in
Eq.~(\ref{eq:stabpar}), and different asymmetry parameters as
functions of $L$.  The theory of Jeans predicts that for our value of
$n=1.5$ the asymmetry $A_{12}$, Eq.~(\ref{eq:18}), in the rotation
plane, should vanish in the continuum limit $N\to\infty$.  In our
simulations it is not exactly zero, but we believe that the small
values we obtain are consistent with zero within statistical
fluctuations.

The asymmetry $A_{(12)3}$, Eq.~(\ref{eq:A123def}), is most directly
comparable to predictions from the continuum theory.  In this case,
unfortunately, we have no theoretical prediction to compare with.  We
have plotted it as a function of $L/L_{\textrm{max}}$, where
$L_{\textrm{max}}$ is the maximum angular momentum, for $N=25$, 50,
150, 400, and 700.  There is a clear tendency that it decreases with
increasing $N$.  Again this may be because the statistical
fluctuations decrease.

With 400 particles we have plotted the average energy per particle
pair, and the ratio between rotational and gravitational energy, as
functions of $L$.  These energy quantities, as well as the asymmetry,
are all fitted very well by polynomials in $L$ of low degree.

In order to illustrate the Jeans effect we have plotted, for 400 and
700 particles, the distribution of particles projected on the vertical
$xz$-plane.  We have also plotted equipotential lines.  The plots are
made with the maximal value of the angular momentum, and also for one
smaller value.  These plots show how the system develops a sharp edge
at the equator, by filling its Roche lobe, when the angular momentum
approaches its maximum value.

\section{General theory}
\label{sec:2}

\subsection{Simulating a polytrope}

The polytropic equation of state gives the gas pressure as
\be
P=K'n^{\gamma}\;,
\ee
where $K'$ is a constant, $n=N/{\cal V}$ is the number density of
particles, $N$ is the number of particles in a volume ${\cal V}$, and
$\gamma$ is the adiabatic index.  A nonrelativistic monatomic gas has
$\gamma=5/3$.  This is the equation of state for the degenerate
electron gas inside a white dwarf star, and it is also the equation of
state in the convection zone inside the Sun.

If the gas expands adiabatically the change in its internal energy $U$
is given by the equation
\be
\rmd U=-P\,\rmd{\cal V}
=-\frac{K'N^{\gamma}}{{\cal V}^{\gamma}}\,\rmd{\cal V}\;,
\ee
which can be integrated to give the internal energy as a function of
the volume,
\be
U=U_0+\frac{K'N^{\gamma}}{(\gamma-1){\cal V}^{\gamma-1}}\;.
\ee
The internal energy is the kinetic energy of the gas molecules, plus
the rotational energy if the molecules rotate.  We now observe that
the same volume dependence of the internal energy is obtained if we
postulate a gas of stationary point particles having a repulsive
potential between a pair of particles at a distance $r$ of the form
\be
\label{eq:Urep}
U_{\rm rep}=\frac{C}{r^{3(\gamma-1)}}\;,
\ee
where $C$ is a constant.  With $\gamma=5/3$ this is an inverse square
potential, $U_{\rm rep}=C/r^2$.

In a gas with $N$ particles we introduce the following repulsive
potential energy between particle pairs, simulating the kinetic energy
of the gas particles,
\be
U_{\rm rep}
=\sum_{i=1}^{N-1}\sum_{j=i+1}^N
\frac{1}{|\vec{r}_i-\vec{r}_j|^2}\;.
\ee
We choose the mass of the identical gas particles as our unit of mass.
Then we choose units of length and time such that $C=1$.  In the
following we will also set the gravitational constant equal to one.

For a fixed number of particles this repulsion energy scales with the
volume in the same way as the kinetic energy of a gas satisfying the
polytropic equation of state with adiabatic index $\gamma=5/3$.
However, as a function of the particle number $N$ it scales
quadratically, as $N(N-1)/2$, whereas the kinetic energy scales
linearly, as $N$.  Thus it is only approximately true that the
pairwise potential energy we introduce may represent kinetic energy.

\subsection{Simulating rotation}
\label{sec:2p2}

We consider a system of $N$ identical particles of unit mass, rotating
as a rigid body about an axis which we take to be the $z$ axis.  We
describe it in a rotating coordinate system where the particle
positions are fixed, thus there are no Coriolis forces.  It has
angular momentum $L=I\Omega$, where $\Omega$ is the angular velocity,
and $I$ is the moment of inertia about the $z$ axis,
\be
I=\sum_{i=1}^N(x_i^{\;2}+y_i^{\;2})\;.
\ee
Conservation laws require that the rotation axis goes through the
centre of mass, hence we impose the constraints
\be
\label{eq:7}
\sum_{i=1}^N\vec{r}_i=0\;.
\ee
The rotational energy is
\be
\label{eq:Erot}
E_{\rm rot}=\frac{1}{2}\,I\Omega^2=\frac{L^2}{2I}\;.
\ee
The other contributions to the energy are the positive repulsion
energy $U_{\rm rep}$ and the negative gravitational potential energy
\be
V_{\rm grav}
=-\sum_{i=1}^{N-1}\sum_{j=i+1}^N
\frac{1}{|\vec{r}_i-\vec{r}_j|}\;.
\ee

When we simulate the rotating system numerically it is physically
meaningful to specify the rotation by fixing the angular momentum $L$,
rather than the angular velocity $\Omega$, since $L$ is the conserved
quantity.  Given the value of $L$ we want to find a configuration
which minimizes the total energy
\be
\label{eq:10}
E=U_{\rm rep}+V_{\rm grav}+E_{\rm rot}
=\sum_{i=1}^{N-1}\sum_{j=i+1}^N\left(
 \frac{1}{|\vec{r}_i-\vec{r}_j|^2}
-\frac{1}{|\vec{r}_i-\vec{r}_j|}\right)
+\frac{L^2}{2I}\;.
\ee

In particular, the energy should be a minimum under the scaling
transformation
\be
\vec{r}_i\to\vec{r}_i\,\!'=\alpha\vec{r}_i
\ee
for $i=1,2,\ldots,N$.  The energy scales as
\be
E\to
E'
=U'_{\rm rep}+V'_{\rm grav}+E'_{\rm rot}
=\frac{U_{\rm rep}}{\alpha^2}
+\frac{V_{\rm grav}}{\alpha}
+\frac{E_{\rm rot}}{\alpha^2}\;.
\ee
The minimum with respect to $\alpha$ is given by the equation
\be
\label{eq:virthproof}
0=\dd{E'}{\alpha}
=-\frac{2U_{\rm rep}}{\alpha^3}
-\frac{V_{\rm grav}}{\alpha^2}
-\frac{2E_{\rm rot}}{\alpha^3}
=-\frac{1}{\alpha}\,(2U'_{\rm rep}+V'_{\rm grav}+2E'_{\rm rot})
\;.
\ee
This shows that after the minimization of the energy $E$ the virial
theorem holds in the form
\be
\label{eq:virth}
2U_{\rm rep}+V_{\rm grav}+2E_{\rm rot}=0\;.
\ee
The virial theorem implies that the total energy in the minimum energy
configuration is half the gravitational energy,
\be
E=U_{\rm rep}+V_{\rm grav}+E_{\rm rot}
=\frac{1}{2}\,V_{\rm grav}\;.
\ee
It also implies for the ratio between the rotational energy and the
gravitational potential energy that
\be
\label{eq:16}
W=\frac{E_{\rm rot}}{|V_{\rm grav}|}
=\frac{1}{2}-\frac{U_{\rm rep}}{|V_{\rm grav}|}
\leq\frac{1}{2}\;.
\ee

\subsection{The potential of a test particle}
\label{sec:2p2a}

If we add another particle of a small mass $\delta m$ at position
$\vec{r}$, without changing the angular momentum, then the change in
energy is
\be
\delta E=\delta m\,\phi(\vec{r})\;.
\ee
This defines the potential $\phi(\vec{r})$.  The rotational energy,
Eq.~(\ref{eq:Erot}), changes because the moment of inertia $I$ and the
angular velocity $\Omega$ change.  The centre of mass is shifted from
$\vec{R}=0$ to
\be
\delta\vec{R}=\frac{\delta m\,\vec{r}}{M+\delta m}\;,
\ee
where $M=N$ is the total mass before addition of the extra particle.
The moment of inertia changes to
\be
I+\delta I
=\sum_{i=1}^N((x_i-\delta X)^2+(y_i-\delta Y)^2)
+\delta m\,((x-X)^2+(y-Y)^2)\;.
\ee
To first order in $\delta m$ we have that
\be
\delta I
=\delta m\,(x^2+y^2)\;.
\ee
To the same order, the change in rotational energy is
\be
\delta E_{\rm rot}=-\frac{L^2}{2I^2}\,\delta I
=\delta m\,\phi_{\rm rot}(\vec{r})\;,
\ee
where $\phi_{\rm rot}$ is the centrifugal potential,
\be
\phi_{\rm rot}(\vec{r})=-\frac{1}{2}\,\Omega^2(x^2+y^2)\;.
\ee
Making the (somewhat arbitrary) assumption that the repulsive
potential is also proportional to $\delta m$ we get the following
expression for the total potential,
\be
\label{eq:phi}
\phi(\vec{r})
=\sum_{i=1}^N\left(
 \frac{1}{|\vec{r}-\vec{r}_i|^2}
-\frac{1}{|\vec{r}-\vec{r}_i|}\right)
-\frac{1}{2}\,\Omega^2(x^2+y^2)\;.
\ee
This is the potential which is plotted in the
Figures~\ref{fig:12and14d} and~\ref{fig:12and14dd}.

These plots of the potential illustrate the stability of the
configurations.  A configuration is stable if it is surrounded by a
potential barrier that prevents particles from escaping.  When its
angular momentum is increased until it becomes unstable, particles
will first escape through one or more Lagrange points, which are
saddle points of the potential.

The equipotential surface through the lowest Lagrange point is the
boundary of what we may call a Roche lobe.  The configuration becomes
unstable when it fills its Roche lobe.  By definition, a saddle point
is a point where equipotential lines cross, hence the boundary of the
Roche lobe must have a cusp, or a sharp edge, there.  Therefore an
axisymmetric cloud of particles filling its Roche lobe must have a
sharp edge, like a discus, as described by Jeans.

\subsection{Stability}
\label{sec:2p3}

For a given value of $L$, any local or global minimum of the energy
$E$ as given in Eq.~(\ref{eq:10}) is a stable equilibrium
configuration.  It is an important observation that $E$ has always at
least one global minimum, because there is an obvious lower bound
$E\geq -N(N-1)/8$.  In fact, the potential energy of two particles at
a distance $d$,
\be
E_2=\frac{1}{d^2}-\frac{1}{d}\;,
\ee
has a minimum $E_2=-1/4$ at $d=2$.  Thus, for any value of $L$ there
exists at least one stable equilibrium configuration.

We will now see how to test numerically for the stability of an
equilibrium configuration.

It is convenient to write the coordinates of the $N$ particles as
\be
\bfn{u}^T
=(u_1,u_2,\ldots,u_{3N})
=(x_1,y_1,z_1,x_2,y_2,\ldots,z_N)\;.
\ee
The superscript $T$ denotes the transpose, thus $\bfn{u}$ is a column
vector.  In order to impose the constraints of Eq.~(\ref{eq:7}) we
subtract the centre of mass position
\be
\vec{R}=\frac{1}{N}\sum_{i=1}^N\vec{r}_i\;.
\ee
The transformation from $\vec{r}_i$ to
$\vec{r}_i\,\!'=\vec{r}_i-\vec{R}$ may be written as
\be
\bfn{u}'=\bfn{Pu}
\ee
where $\bfn{P}$ is a $(3N)\times(3N)$ matrix.  We write $\bfn{I}_n$
for the $n\times n$ identity matrix, then
\be
\bfn{P}=
\bfn{I}_{3N}-\frac{1}{N}
\begin{pmatrix}
\bfn{I}_3 & \bfn{I}_3 & \hdots & \bfn{I}_3\\
\bfn{I}_3 & \bfn{I}_3 & \hdots & \bfn{I}_3\\
\vdots & \vdots & \ddots & \vdots\\
\bfn{I}_3 & \bfn{I}_3 & \hdots & \bfn{I}_3
\end{pmatrix}.
\ee
$\bfn{P}$ is an orthogonal projection, which means that
$\bfn{P}^2=\bfn{P}$ and $\bfn{P}^T=\bfn{P}$.

Consider a perturbation $\bfn{u}\to\bfn{u}+\epsilon\bfn{v}$ where
$\epsilon$ is a small parameter.  The perturbation is physically
meaningful if $\bfn{v}=\bfn{Pw}$ for some vector $\bfn{w}$, so that
the centre of mass is not moved away from the rotation axis.

The condition for stable equilibrium is that the energy $E$ is minimal
at $\epsilon=0$ for any direction vector $\bfn{v}=\bfn{Pw}$.  This
means that the first derivative with respect to $\epsilon$ must
vanish, and the second derivative must be non-negative.  The first
derivative must vanish also for an unphysical perturbation moving the
centre of mass away from the rotation axis, hence all the partial
derivatives must vanish,
\be
\frac{\partial E}{\partial u_i}=0
\qquad\mbox{for}\quad i=1,2,\ldots,3N\;.
\ee
The condition of non-negative second derivative is that the matrix
\be
\bfn{M}=\bfn{PDP}\;,
\ee
where $\bfn{D}$ is the so called 
Hessian matrix of second derivatives,
\be
\label{eq:17}
D_{ij}=\frac{\partial^2E}{\partial{u_{i}}\partial{u_{j}}}
\;,
\ee
must have only non-negative eigenvalues.  Our numerical computations
give the eigenvalues in increasing order,
\be
\lambda_1\leq\lambda_2\leq\lambda_3\leq\ldots\;.
\ee
Since we have set all the particle masses equal to one, the physical
interpretation of the eigenvalues is that each $\sqrt{\lambda_i}$ is a
frequency of vibration if we introduce equations of motion.  The
corresponding eigenvector describes the mode of vibration.

Four eigenvalues vanish identically, because they correspond to
overall translations in three directions and an overall rotation about
the rotation axis.  If the configuration is stable, then
$\lambda_1=\cdots=\lambda_4=0$ and $\lambda_5$ is the smallest
positive eigenvalue, corresponding to the least stable mode of
vibration.  If the configuration has $k$ unstable modes, $k\geq 1$,
then $\lambda_1\leq\cdots\leq\lambda_k<0$ and
$\lambda_{k+1}=\cdots\lambda_{k+4}=0$.  Therefore we define a
stability parameter
\be
\label{eq:stabpar}
\sigma = \lambda_1+\lambda_2+\lambda_3+\lambda_4+\lambda_5\;,
\ee
which is positive or negative depending on whether the configuration
is stable or unstable~\cite{HM}.

A special case is when the configuration does not rotate.  Then six
eigenvalues, corresponding to three translations and three rotations,
vanish identically.  In that case it is still true that the
configuration is stable when $\sigma\geq 0$.

\subsection{The asymmetry parameters}
\label{subsec:Asymmetry}

In order to compare the shapes of our simulated rotating configurations 
with the shapes of rotating liquid drops we introduce a matrix
\be
\bfn{J}=
\sum_{i=1}^N
\begin{pmatrix}
 x_ix_i & x_iy_i & x_iz_i \\
 y_ix_i & y_iy_i & y_iz_i \\
 z_ix_i & z_iy_i & z_iz_i  
\end{pmatrix},
\ee
the elements of which are central moments of the mass distribution.
The eigenvectors of this matrix are the principal axes of the body,
and we choose them as our $x,y,z$ axes.  The corresponding eigenvalues
$\beta_1\geq\beta_2\geq\beta_3>0$ are the moments along the principal
axes.  We order them in decreasing order, then the rotation axis will
always be the $z$ axis, since the rotation flattens the body.  We
define asymmetry parameters
\be
\label{eq:18}
A_{ij} = \frac{\beta_i-\beta_j}{\beta_i+\beta_j}\;.
\ee 
These definitions imply that $0\leq A_{ij}<1$ when $i<j$.
The three different asymmetries satisfy the relations
\be
\label{eq:asymrelations}
A_{12}=\frac{A_{13}-A_{23}}{1-A_{13}A_{23}}\;,\qquad
A_{13}=\frac{A_{12}+A_{23}}{1+A_{12}A_{23}}\;,\qquad
A_{23}=\frac{A_{13}-A_{12}}{1-A_{13}A_{12}}\;.
\ee
Since  $A_{12}\geq 0$ and $A_{23}\geq 0$
it follows that $A_{13}\geq A_{23}$ and $A_{13}\geq A_{12}$.  In the
examples where we compute the asymmetries, we always have
$A_{12}\approx 0$ and $A_{23}\approx A_{13}$.

The asymmetries computed in our discrete model may be compared with
similar quantities computed in continuum models, such as polytropes
and incompressible fluids.  In the continuum case there may be perfect
rotational symmetry in the $xy$-plane.  In the corresponding discrete
model, the rotational symmetry in the plane will be only approximate,
due to statistical fluctuations if not for other reasons, so that the
eigenvalues $\beta_1$ and $\beta_2$ will differ slightly.  For the
comparison we should then introduce the mean value
$\beta_{12}=(\beta_1+\beta_2)/2$ and define
\be
\label{eq:A123def}
A_{(12)3}=\frac{\beta_{12}-\beta_3}{\beta_{12}+\beta_3}\;.
\ee

According to the Jeans effect, as we have called it, for a polytrope
with angular momentum less than maximal there is perfect rotational
symmetry in the plane.  As the angular momentum approaches its maximal
value the mechanism for instablility is that particles escape from the
equator.  With a finite number of particles we see the same mechanism
for instability at maximal angular momentum, and furthermore the
escape of one particle necessarily results in a nonzero asymmetry
$A_{12}$.  Since it is impossible to lose less than one particle, it
is reasonable to call this process a statistical fluctuation.  This
may explain the appearance of Fig.~\ref{fig:A12400}, where it is seen
that $A_{12}$ increases sharply just before the instability sets in.

In the classical theory of rotating incompressible liquid bodies bound
by gravitation the equilibrium shapes are ellipsoids.  Denote the half
axes of the ellipsoid by
\be
a_1\geq a_2\geq a_3\;.
\ee
The eccentricities are defined as
\be
e_{ij}=\sqrt{1-\frac{a_j^{\,2}}{a_i^{\,2}}}
\ee
when $a_i>a_j$.  The three different eccentricities satisfy the
relation
\be
1-e_{13}^{\,2}=(1-e_{12}^{\,2})(1-e_{23}^{\,2})\;.
\ee
The second moments of the ellipsoid are proportional
to $a_1^{\,2},a_2^{\,2},a_3^{\,2}$, hence the asymmetry parameters of
the ellipsoid, with $i<j$, are
\be
\label{eq:18a}
A_{ij} = \frac{a_i^{\,2}-a_j^{\,2}}{a_i^{\,2}+a_j^{\,2}}
 =\frac{e_{ij}^{\,2}}{2-e_{ij}^{\,2}}\;.
\ee 
The other way around, the eccentricities are expressed in terms of the
asymmetries as
\be
e_{ij}=\sqrt{\frac{2A_{ij}}{1+A_{ij}}}\;.
\ee

\section{Configurations of many particles without rotation}
\label{sec:3}

In this section, and the next, we present results from numerical
studies of the minimum energy configurations without rotation.  We
consider first the case when the number of particles, $N$, is large.
In the next section we will present plots showing what the minimum
energy configurations look like for small $N$.

The energy $E$ becomes more negative, and the geometrical size
increases somewhat, although not very much, as $N$ increases.  In our
calculations we find approximate scaling laws for the dependence on
$N$ of the energy and size.  We argue that for large $N$ the density
distribution of particles approaches that of a nonrotating
gravitationally bound polytrope of polytropic index $n=3/2$, as
described by the Lane--Emden equation.  Since this is the theoretical
model that we compare our numerical results against, we begin by
describing it briefly.

\subsection{The Lane--Emden equation}

In Appendix~\ref{app:laneemden} we discuss the numerical treatment of
the Lane--Emden equation.  We want to compare some numbers computed
from the configurations we generate in our model, with the
corresponding numbers computed from the Lane--Emden density
distribution.  These can be expressed as integrals that most often
have to be computed numerically.  We use the Monte Carlo method to
integrate numerically.  That is, we generate random configurations of
a large number of noninteracting particles from the given density
distribution, as described in the appendix.

We define a function $f(u)$ satisfying the Lane--Emden equation on the
interval $0\leq u\leq 1$, with $f'(0)=0$ and $f(1)=0$.  The particle
number density, or mass density if each particle has unit mass, at a
radius $r=bu$ where $b$ is a scaling factor, is
\be
\rho(r)=\beta\,(f(u))^{3/2}\;.
\ee
Here $\beta\,(f(0))^{3/2}$ is the central density.  The fraction of
the mass inside the radius $r$ is the function $F(u)$ defined in
Eq.~(\ref{eq:Fu}).

There is a basic difference between the Lane--Emden model and our
model with a pair interaction consisting of a gravitational attraction
of long range, and a repulsion of shorter range introduced
artificially to prevent collapse.  In the Lane--Emden case there is no
repulsion between particles, hence in the Monte Carlo process two
particles can come arbitrarily close to each other.  In our model,
there is a lower limit to the distance between two particles when $N$
is fixed, because it costs energy to bring them close together.

What happens in our model when we increase $N$, is essentially that we
fill up with more particles inside a volume that does not increase
very much.  We will argue next that the volume must stay nearly
constant because of the virial theorem.  It means that the smallest
interparticle distance must decrease roughly as $N^{-1/3}$, in our
model.  This explains why $N^{-1/3}$ is a useful expansion parameter
when we study how various quantities vary with $N$ in the limit of
large $N$, a fact that we discovered by trial and error.
Unfortunately, $N^{-1/3}$ goes rather slowly to zero as $N\to\infty$.
This means that in practice we can never quite neglect finite size
effects.

\subsection{Energy as a function of $N$}

\begin{figure}[h]
  \begin{center}
\fbox {\includegraphics[width=7.48cm]{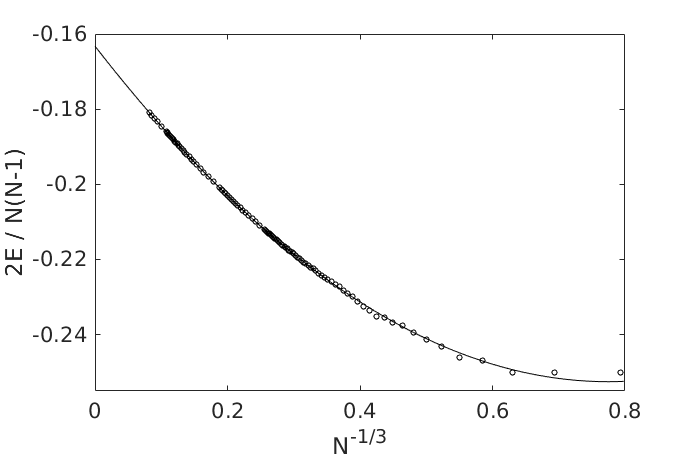}}
\fbox {\includegraphics[width=7.48cm]{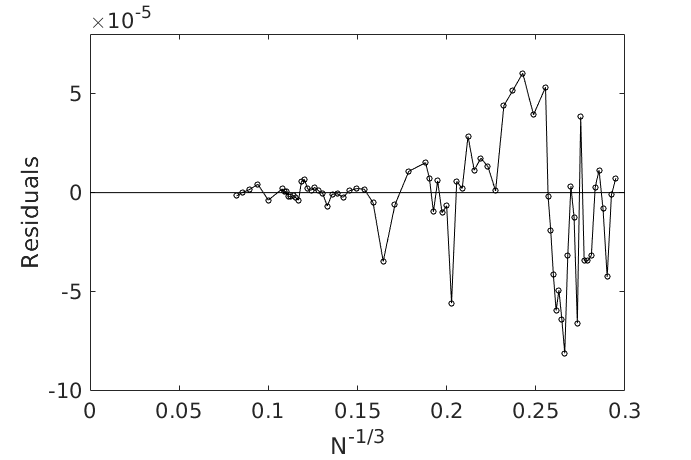}}
\caption{\label{fig:forhE} The average energy per particle pair,
  $\langle E_2\rangle$, as a function of $N^{-1/3}$.  The line is the
  fitted cubic polynomial in $N^{-1/3}$ given in
  Eq.~(\protect{\ref{eq:EphN2}}).  The panel to the right shows the
  residuals of the fit for $N\geq 39$.  It shows magic numbers at
  $N=250$, $N=120$, and $N=53$.}
\end{center}
\end{figure}

The number of particle pairs is $N(N-1)/2$, and when the total energy
is $E$ the average energy per particle pair is
\be
\langle E_2\rangle=\frac{2E}{N(N-1)}\;.
\ee
The computed minimal values of $\langle E_2\rangle$ for some values of
$N$ are tabulated in Table~\ref{tab:Grav}.  These and many more values
are plotted in Fig.~\ref{fig:forhE}.  By the virial theorem, after
minimization of the energy the gravitational energy is
$V_{\mathrm{grav}}=2E$.

\begin{table}[h]
\begin{center}
\begin{tabular}{rlrrl}
\multicolumn{1}{c}{$N$} &
\multicolumn{1}{c}{$\langle E_2\rangle$} &&
\multicolumn{1}{c}{$N$} &
\multicolumn{1}{c}{$\langle E_2\rangle$}\\ 
\cline{1-2}
\cline{4-5}\\
$ 5$ &   $-0.246\,917\,186$      &&    $26$  &  $-0.223\,736\,991$\\
$ 6$ &   $-0.246\,187\,269$      &&    $53$  &  $-0.213\,726\,499^*$\\
$ 7$ &   $-0.243\,083\,560$      &&   $120$  &  $-0.203\,514\,569$\\ 
$ 8$ &   $-0.241\,178\,876$      &&   $225$  &  $-0.196\,783\,556$\\ 
$ 9$ &   $-0.239\,491\,029$      &&   $400$  &  $-0.191\,451\,634$\\ 
$10$ &   $-0.237\,629\,424$      &&   $575$  &  $-0.188\,491\,629$\\ 
$10$ &   $-0.237\,600\,140^*$    &&   $800$  &  $-0.186\,053\,272$\\ 
$11$ &   $-0.236\,721\,422^*$    &&  $1000$  &  $-0.184\,531\,766$\\  
$12$ &   $-0.235\,486\,032^*$    &&  $1200$  &  $-0.183\,346\,25\phantom{0}$\\  
$13$ &   $-0.235\,117\,408^*$    &&  $1400$  &  $-0.182\,400\,59\phantom{0}$\\  
$14$ &   $-0.233\,447\,602^*$    &&  $1600$  &  $-0.181\,614\,28\phantom{0}$\\ 
$15$ &   $-0.232\,410\,680^*$    &&  $1800$  &  $-0.180\,946\,32\phantom{0}$
\end{tabular}
\caption{\label{tab:Grav} The minimal value of the average energy per
  particle pair, $\langle E_2\rangle$, for $N$ particles without
  rotation.  The asterisks mark configurations for small $N$ with a
  central particle (not exactly at the centre for $N=12$ and $N=14$).}
\end{center}
\end{table}

Since the minimum energy of one particle pair is $-1/4$ at a distance
of $d=2$, it is clear that $\langle E_2\rangle\geq -1/4$.  Obviously,
this is a strict inequality for $N>4$, since not all particle pairs can
have the optimal distance $d=2$.

The figure shows a slow increase of $\langle E_2\rangle$ with $N$.  It
also shows the existence of ``magic numbers'', where the energy is
exceptionally low because the particles are arranged in configurations
of high symmetry.  Some magic numbers are 4 (a regular tetrahedron), 6
(a regular octahedron), 13 (a regular icosahedron with a central
particle), 26, and 53.  In Section~\ref{sec:4} we will show the
configurations with $N=26$ and $N=53$.

The average energy per particle pair is well fitted over the whole
range, down to $N=2$, by a cubic polynomial in $N^{-1/3}$,
\be
\label{eq:EphN2}
E_2^{\textrm{fit}}=a_0+a_1N^{-1/3}+a_2N^{-2/3}+a_3N^{-1}\;,
\ee
with the following values of the coefficients,
\be
a_0=-0.1632304\;,\quad
a_1=-0.2270268\;,\quad
a_2=0.1399513\;,\quad
a_3 =0.0055762\;.
\ee
It must be noted that we have done a weighted least squares fit with
weights $N^2$, because the important point is to obtain a good fit to
the energies at the largest values of $N$.  Nevertheless the fit is
reasonably good all the way down to $N=2$.

The fit gives us a limiting value as $N\to\infty$ which is
\be
\langle E_2\rangle_{\infty}=a_0=-0.1632304\;.
\ee
This can be compared to the number computed by Monte Carlo from the
Lane--Emden distribution, as given in the appendix,
\be
\langle E_2\rangle_{\textrm{LE}}=-0.16253\pm 0.00015\;.
\ee

\begin{figure}
\begin{center}
\fbox {\includegraphics[width=8cm]{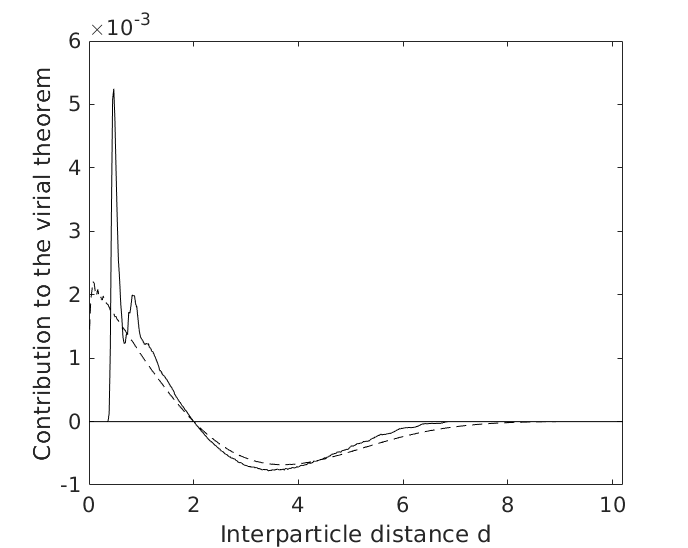}}
\caption{\label{fig:bin} This plot shows how different values of the
  interparticle distance $d$ contribute to the virial theorem,
  Eq.~(\protect{\ref{eq:virth1}}).  The contribution is positive for
  $d<2$ and negative for $d>2$.  The fulldrawn, peaked curve is for
  1800 particles in our model.  The dashed, smoother curve is for
  10\,000 independent points drawn at random from the Lane--Emden
  density distribution.  This curve should be essentially independent
  of the number of points.  For the sake of the comparison, both
  curves are divided by the number of particle pairs.}
\end{center}
\end{figure}

We understand the existence of this limit as a consequence of the
virial theorem.  The energy of one configuration is
\be
\label{eq:Esumpairs}
E=\sum_{\mbox{all pairs}}\left(\frac{1}{d^2}-\frac{1}{d}\right),
\ee
where $d$ is the distance between the particles of one pair.  After
$E$ has been minimized, the virial theorem must hold, which says that
\be
\label{eq:virth1}
\sum\left(\frac{2}{d^2}-\frac{1}{d}\right)=0\;.
\ee
In the last sum, the quantity in parenthesis is positive for $d<2$ and
negative for $d>2$.  The sum can be zero only if the interparticle
distances $d$ are distributed equally below and above $d=2$, as shown
in Fig.~\ref{fig:bin}, where we have plotted the contributions to the
sum from the different values of $d$.  This balance fixes the
geometrical size of the cloud of particles.  As we increase $N$ the
size must remain nearly constant, and all that happens, roughly
speaking, is that the density of particles increases everywhere in the
same proportion.

\subsection{Finite size effects}

Figure~\ref{fig:bin} demonstrates clearly how a finite system is
different from the continuum limit, in particular because there is a
lower limit to the distance between particles.  In order to understand
the limit $N\to\infty$, we have studied what happens when $N$
increases.

We define $\langle r\rangle$ as the root mean square distance of the
particles from the origin.  We write $d$ for the distance between two
particles, and write $\langle d\rangle$ for the average of $d$ over
all the particle pairs.  The mean values $\langle r\rangle$ and
$\langle d\rangle$ are tabulated in Table~\ref{tab:Nrd} for some
values of $N$.

\begin{table}[h]
\begin{center}
\begin{tabular}{rrrrrrr}
\multicolumn{1}{c}{$N$} &
\multicolumn{1}{c}{$\langle r\rangle$} &
\multicolumn{1}{c}{$\langle d\rangle$} &&
\multicolumn{1}{c}{$N$} &
\multicolumn{1}{c}{$\langle r\rangle$} &
\multicolumn{1}{c}{$\langle d\rangle$}\\
\cline{1-3}
\cline{5-7}\\
   10 & 1.532\,947\,447\,00 & 2.223\,550\,231\,57&&
  500 & 2.408\,216\,555\,94 & 3.195\,863\,786\,49\\
   20 & 1.746\,229\,728\,92 & 2.424\,671\,301\,48&&
  600 & 2.432\,871\,960\,05 & 3.227\,071\,970\,79\\
   30 & 1.852\,246\,091\,36 & 2.535\,394\,051\,13&&
  625 & 2.438\,273\,052\,75 & 3.233\,930\,049\,01\\
   40 & 1.926\,089\,543\,71 & 2.617\,026\,807\,14&&
  700 & 2.452\,576\,067\,99 & 3.252\,214\,640\,82\\
   56 & 2.006\,652\,372\,32 & 2.708\,188\,019\,60&&
  800 & 2.469\,284\,455\,65 & 3.273\,392\,872\,15\\
   70 & 2.059\,629\,841\,53 & 2.769\,512\,183\,49&&
 1000 & 2.495\,284\,325\,54 & 3.306\,661\,605\,26\\
   80 & 2.087\,918\,877\,46 & 2.802\,950\,867\,71&&
 1200 & 2.515\,646\,479\,04 & 3.332\,774\,038\,07\\
  120 & 2.170\,644\,712\,76 & 2.902\,144\,797\,17&&
 1400 & 2.531\,895\,026\,75 & 3.353\,666\,514\,75\\
  350 & 2.356\,905\,374\,94 & 3.131\,156\,455\,70&&
 1600 & 2.545\,429\,383\,56 & 3.371\,090\,162\,89\\
  400 & 2.376\,865\,330\,71 & 3.156\,148\,941\,84&&
 1800 & 2.556\,902\,476\,51 & 3.385\,898\,613\,04\\
  450 & 2.393\,687\,080\,41 & 3.177\,398\,870\,61&&&&
\end{tabular}
\caption{\label{tab:Nrd} The root mean square radius
  $\langle r\rangle$ and the mean distance between particles,
  $\langle d\rangle$, for the minimum energy configuration of $N$
  particles without rotation.}
\end{center}
\end{table}

\begin{figure}
\begin{center}
\fbox {\includegraphics[width=7.48cm]{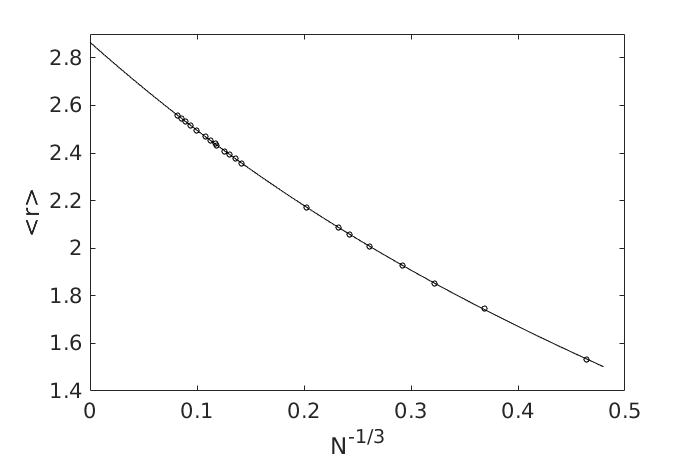}}
\fbox {\includegraphics[width=7.48cm]{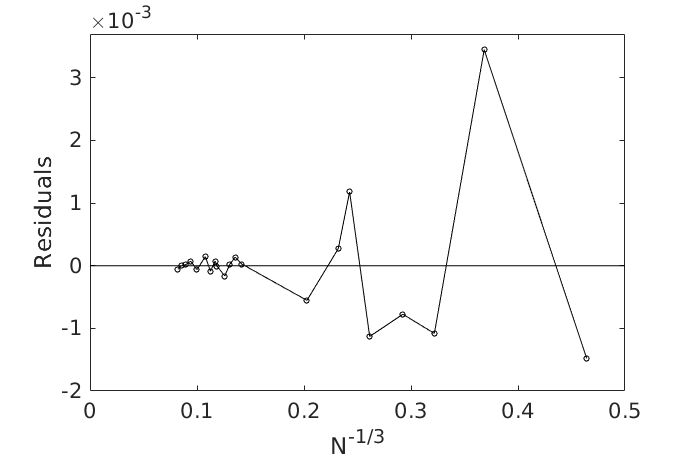}}
\caption{\label{fig:rrms}The root mean square radius
  $\langle r\,\rangle$ as a function of $N^{-1/3}$.  The line is the
  cubic polynomial $r_m$ given in Eq.~(\ref{eq:rm}).  The panel to the
  right shows the residuals of the fit, $\langle r\rangle-r_m$.}
\end{center}
\end{figure}

\begin{figure}
\begin{center}
\fbox {\includegraphics[width=7.48cm]{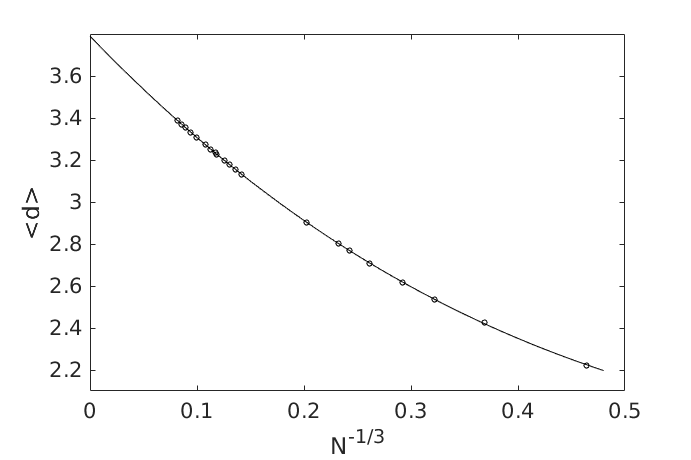}}
\fbox {\includegraphics[width=7.48cm]{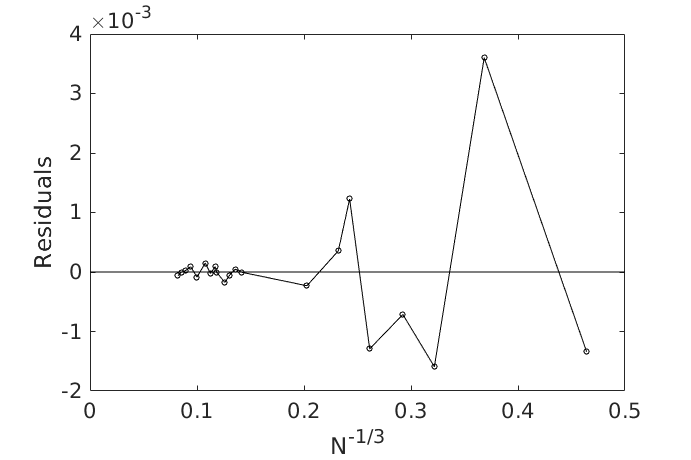}}
\caption{\label{fig:dm}The mean interparticle distance
  $\langle d\,\rangle$ as a function of $N^{-1/3}$.  The line is the
  cubic polynomial $d_m$ given in Eq.~(\ref{eq:dm}).  The panel to the
  right shows the residuals $\langle d\rangle-d_m$.}
\end{center}
\end{figure}

The tabulated data are plotted in the Figures~\ref{fig:rrms}
and~\ref{fig:dm}.  The curves in the figures are cubic polynomials in
$N^{-1/3}$,
\be
\label{eq:rm}
r_m=b_0+b_1N^{-1/3}+b_2N^{-2/3}+b_3N^{-1}\;,
\ee
and
\be
\label{eq:dm}
d_m=c_0+c_1N^{-1/3}+c_2N^{-2/3}+c_3N^{-1}\;,
\ee
with the following fitted values for the coefficients,
\be
b_0 =  2.86513\;,\quad
b_1 = -3.99618\;,\quad
b_2 =  3.13506\;,\quad
b_3 = -1.51269\;,
\ee
and
\be
c_0 =  3.78972\;,\quad
c_1 = -5.30190\;,\quad
c_2 =  4.87658\;,\quad
c_3 = -1.54528\;.
\ee
The least squares fitting is done with weight $N$ for each data point,
emphasizing the high values of $N$.  Also plotted are the residuals
$\langle r\rangle-r_m$ and $\langle d\rangle-d_m$.  They show no
systematic dependence on $N$.  According to the fits there exist
limiting values as $N\to\infty$,
\be
\langle r\rangle_{\infty}=b_0=2.86513\;,\qquad
\langle d\rangle_{\infty}=c_0=3.78972\;.
\ee
These can be compared to the numbers computed by Monte Carlo from the
Lane--Emden distribution, as given in the appendix,
\be
\langle r\rangle_{\textrm{LE}}=2.9220\pm 0.0029\;,\qquad
\langle d\rangle_{\textrm{LE}}=3.8432\pm 0.0037\;.
\ee

\subsubsection*{Approximate scaling}

\begin{figure}
\begin{center}
\fbox {\includegraphics[width=7.48cm]{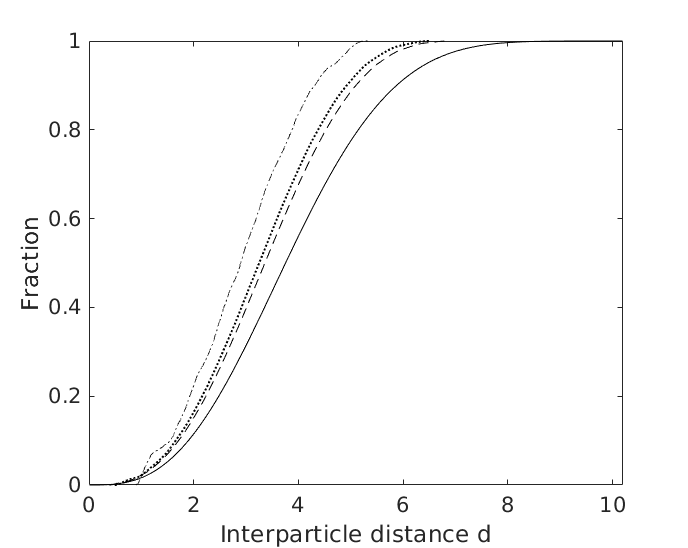}}
\fbox {\includegraphics[width=7.48cm]{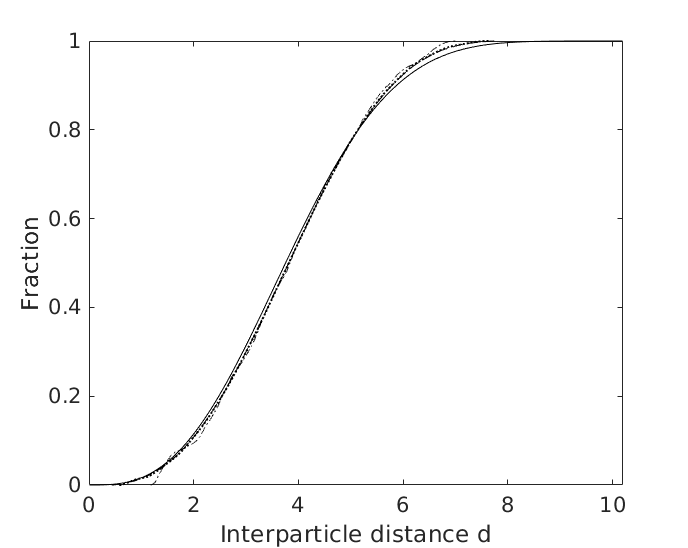}}
\caption{\label{fig:cdd}Cumulative distribution of the interparticle
  distance $d$ for the Lane--Emden distribution (solid line), and our
  model with $N=1800$ (dashed line), $N=800$ (dotted line), and
  $N=120$ (dash-dot line).  The panel to the right shows that all
  curves fall very nearly on top of each other when all distances are
  multiplied by scale factors of $1.1351$ for $N=1800$, $1.1741$ for
  $N=800$, and $1.3243$ for $N=120$, to make the mean values
  $\langle d\rangle$ equal.}
\end{center}
\end{figure}

Figure~\ref{fig:cdd} shows the cumulative distribution of the
interparticle distance $d$ for all the 
particle pairs.  The three cases $N=120$, $N=800$, and $N=1800$ are
shown, together with the presumed $N\to\infty$ limit given by the
Lane--Emden equation.  Remember that one particle pair has its minimum
energy when $d=2$.  Most of the distances are larger, we see for
example from figure~\ref{fig:cdd} that for $N=1800$ we have $d>2$ for
90\,\% of the particle pairs.  In the virial theorem,
Eq.~(\ref{eq:virth1}), the 10\,\% of the distances smaller than two
weigh up for the 90\,\% of the distances larger than two.

All four distributions are seen to have very nearly the same shape.
In fact, they are brought to lie very nearly on top of each other when
the three configurations with $N=120$, $N=800$, and $N=1800$ are
magnified by factors of $1.3243$, $1.1741$, and $1.1351$,
respectively.  These factors are calculated from the mean values in
Table~\ref{tab:Nrd}, and from the mean value $\langle d\rangle=3.8432$
for the Lane--Emden case, see Appendix~\ref{app:laneemden}.

\begin{figure}
\begin{center}
\fbox {\includegraphics[width=7.48cm]{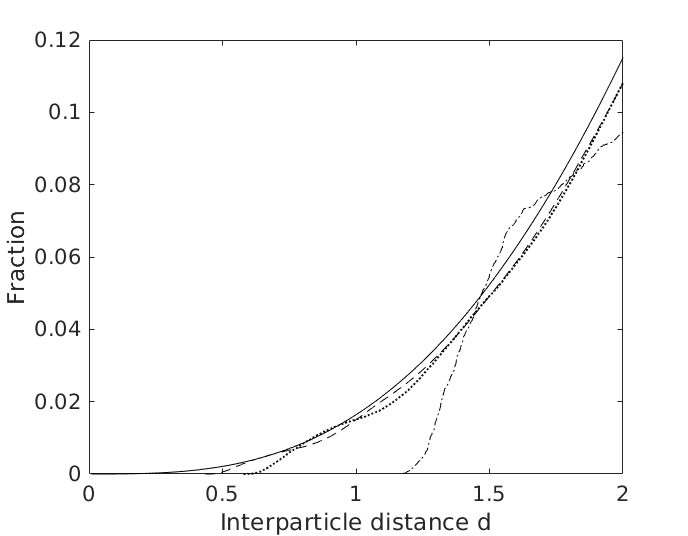}}
\fbox {\includegraphics[width=7.48cm]{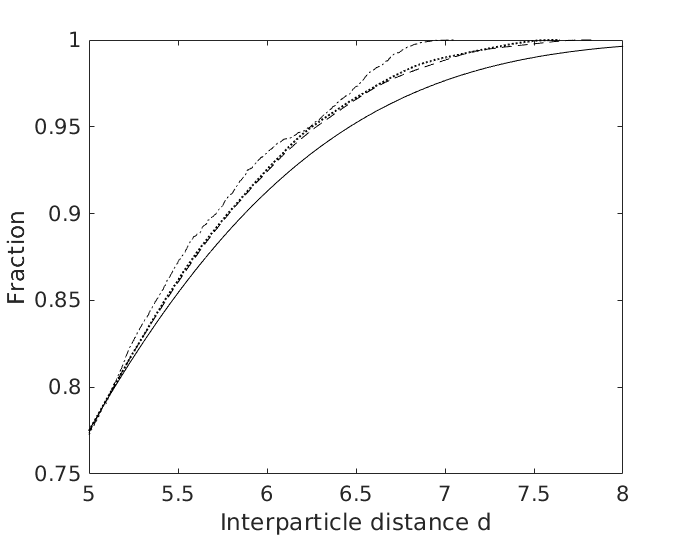}}
\caption{\label{fig:cdd1}The left panel shows details of the lower
  part of the right panel in Fig.~\protect{\ref{fig:cdd}}, and the
  right panel shows the top part of the same figure.  We see that with
  a small number of particles the distribution of interparticle
  distances is cut off both at small and large values.  These cutoffs
  seem to have little effect on the overall density distribution
  computed by smearing out the particles artificially, as shown in
  Fig~\protect{\ref{fig:densityprofile}}.}
\end{center}
\end{figure}

The curves can not have identically the same shape, however, for one
very good reason.  A simple scaling of all the distances $d$ by a
common factor would break the virial theorem, because the two terms in
Eq.~(\ref{eq:virth1}) scale differently.  It follows that the four
curves in Fig.~\ref{fig:cdd} must be visibly different after scaling.
The differences in shapes are seen in Fig.~\ref{fig:cdd1} at both ends
of the curves, while the middle parts of the curves follow each other
closely.

\subsubsection*{Density profiles}

\begin{figure}
\begin{center}
\fbox {\includegraphics[width=7.48cm]{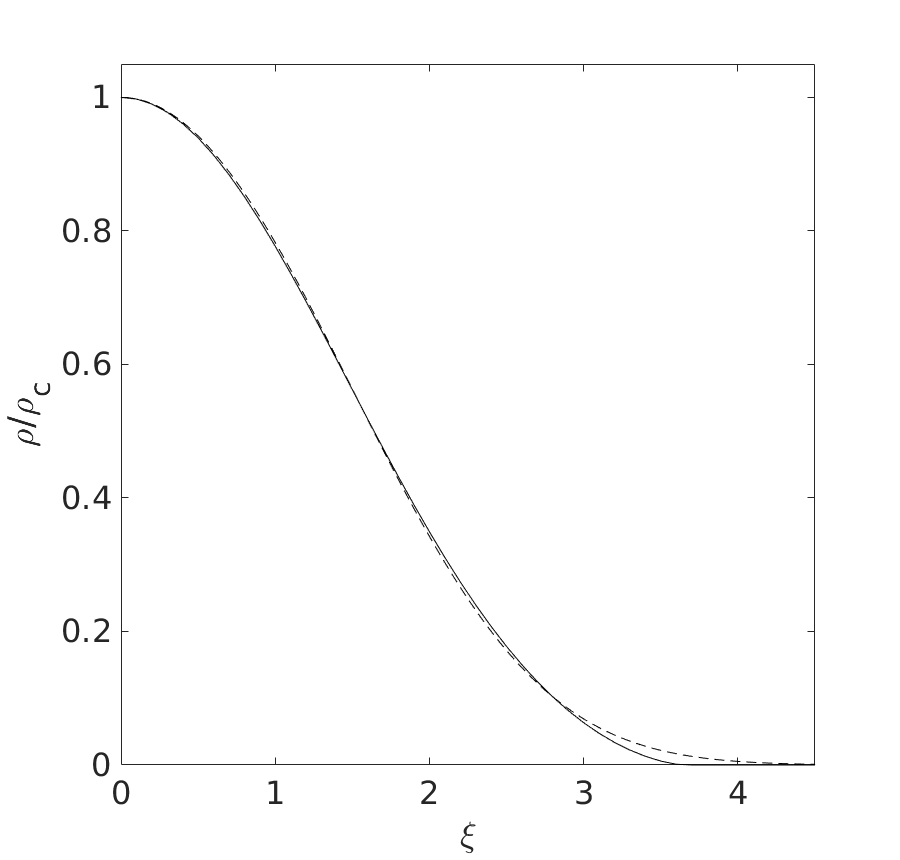}}
\fbox {\includegraphics[width=7.48cm]{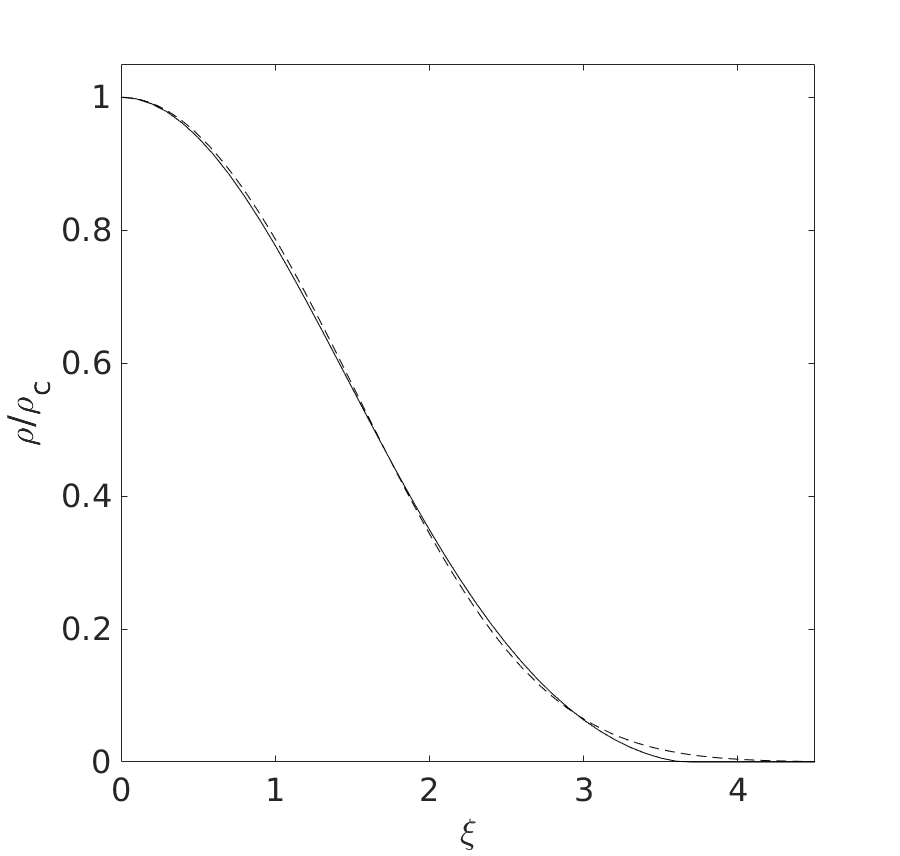}}
\caption{\label{fig:densityprofile} The solid curve in each plot shows
  the dimensionless density $\theta^n$ for $n=3/2$, determined by the
  Lane--Emden equation, as a function of the dimensionless radius
  $\xi$.  The physical radius is $r=a\xi$ where $a$ is a scale factor.
  The dashed curve in the left panel is calculated from our
  nonrotating configuration of 120 point particles, as explained in
  the text, with a scale factor $a=1.485$.  The dashed curve in the
  right panel is for 1800 particles, here $a=1.64$.}
\end{center}                           
\end{figure}

The dimensionless Lane--Emden equation, Eq.~(\ref{eq:LaneEmden}),
describes the density profile of a nonrotating polytropic gas cloud of
given polytropic index $n$.  We take here $n=3/2$.  The density at a
radius $r=a\xi$ is
\be
\rho(r)=\rho_c\,(\theta(\xi))^n\;,
\ee
where $\rho_c$ is the central density and $a$ is a scaling factor.
The solid curves in the two plots in Fig.~\ref{fig:densityprofile}
show the dimensionless density $\theta^n=\rho/\rho_c$ as a function of
the dimensionless radius $\xi$.  The surface of the gas cloud where
$\theta(\xi)=0$, is given by $\xi=\xi_1= 3.653\,754$.  The dashed
curves in these plots show the density profiles we compute for $N=120$
(left) and $N=1800$ (right).  In order to compute a continuous density
profile from a configuration of $N$ point particles, we can imagine
looking through an unfocused telescope, seeing each point particle as
a spherically symmetric gaussian density distribution.  We take the
standard deviation to be $1.36$, comparable to the minimum distance
between points for $N=120$, as Fig.~\ref{fig:cdd1} shows, In order to
make our numerical density profiles coincide with the theoretical
profile for a polytrope, we have to divide $r$, the distance from the
origin to a point where we compute the density, by a scale factor $a$
that is $1.485$ for $N=120$ and $1.64$ for $N=1800$.  The artificial
smearing out of the point particles produces extra tails to the dashed
curves.

These plots of density profiles are again good evidence that the
configurations of finite numbers of point particles that we generate,
can be understood as representing polytropes of index $n=3/2$.

\section{Configurations of few particles without rotation}
\label{sec:4}

In this section we will describe some of the simplest examples of
minimum energy configurations, including the magic numbers $N=26$ and
$N=53$.

\begin{figure}
\begin{center}
\fbox {\includegraphics[width=7.48cm]{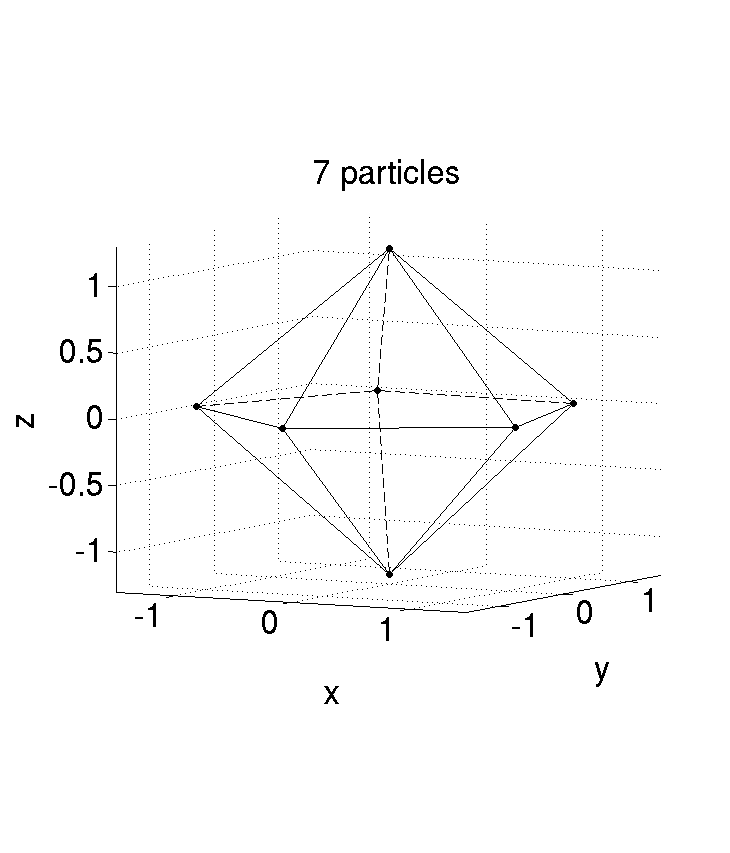}}
\fbox {\includegraphics[width=7.48cm]{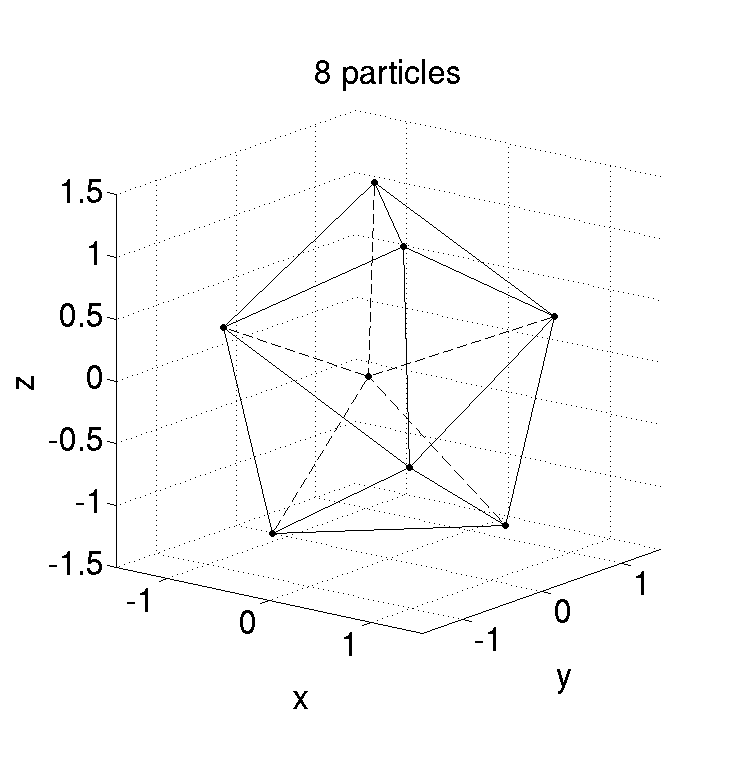}}
\caption{\label{fig:78}Minimum energy configurations.}
\end{center}
\end{figure}

{\bf Five, six, and seven particles.}  Figure~\ref{fig:78} shows the
minimum energy configuration of seven particles.  With five, six, or
seven particles, two of the particles define a symmetry axis, the
$z$-axis in the figure.  The remaining particles form a regular
polygon in the horizontal plane, the $xy$-plane in the figure, an
equilateral triangle when $N=5$, a square when $N=6$, or a regular
pentagon as in Fig.~\ref{fig:78}.  The configuration with six
particles (not shown) has maximal symmetry, since it is a regular
octahedron, symmetric under a group of 48 different rotations and
reflections.

{\bf Eight particles.}  The minimum energy configuration of eight
particles, also shown in Fig.~\ref{fig:78}, can be described as made
out of two paper boats, where one is turned upside down, rotated
$90^{\circ}$ and put on top of the other.  It is a polytope with four
four-fold corners (where four edges meet), and four five-fold corners.
It has a special symmetry which is a simultaneous reflection in the
$xy$-plane and a rotation by $90^{\circ}$.  It also has two vertical
symmetry planes (through the $z$-axis).

\begin{figure}
\begin{center}
\fbox {\includegraphics[width=7.48cm]{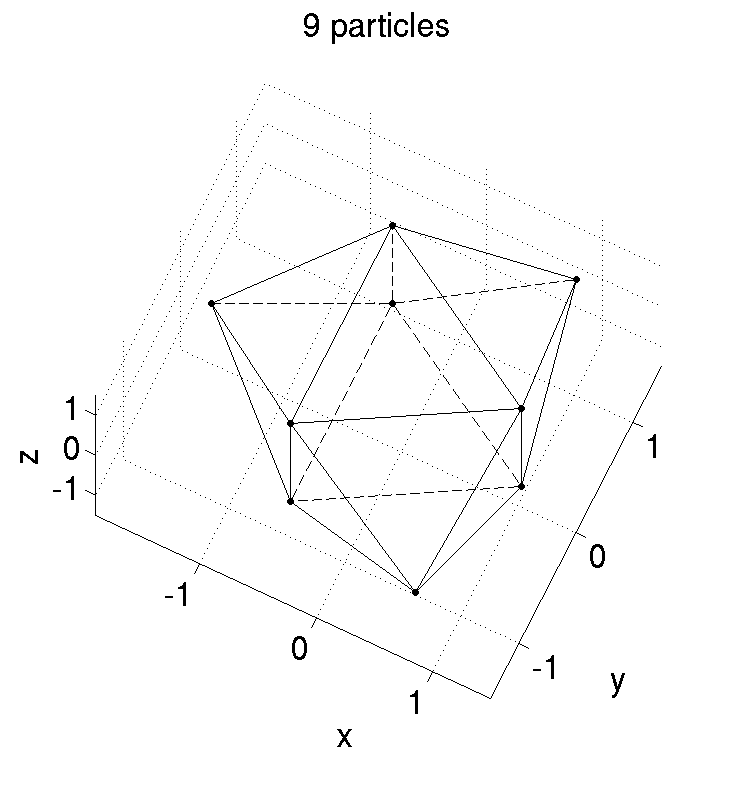}}
\fbox {\includegraphics[width=7.48cm]{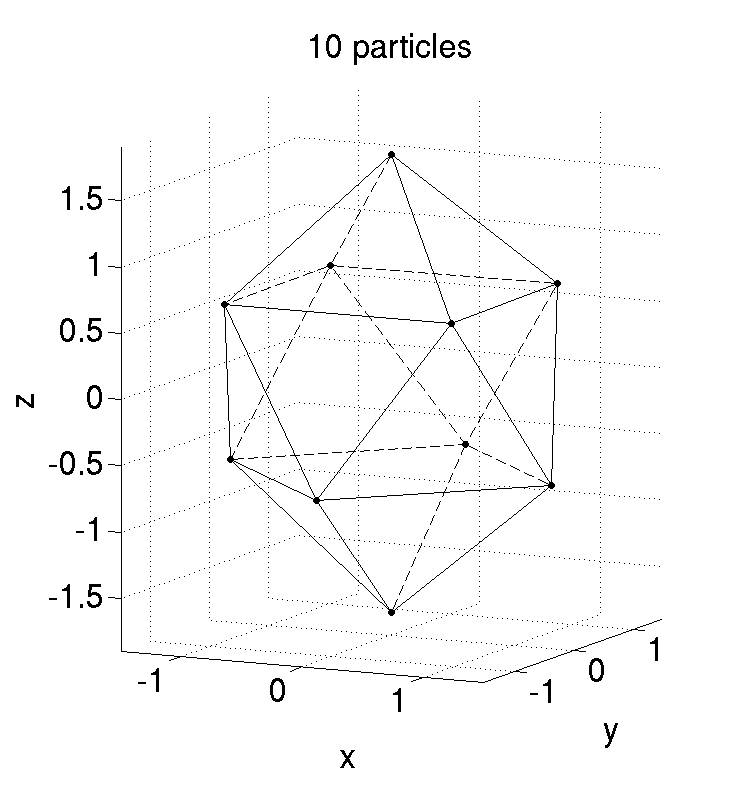}}
\caption{\label{fig:9and10}  Minimum energy configurations.}
\end{center}
\end{figure}

{\bf Nine particles.}  The minimum energy configuration with nine
particles is shown in Figure~\ref{fig:9and10}.  It is a polytope with
three four-fold and six five-fold corners.  It is symmetric under
rotations by $120^{\circ}$ about the $z$ axis.  The horizontal plane
($xy$-plane) and three vertical planes are symmetry planes.

{\bf Ten particles.}  With ten particles there are two widely
different configurations that are nearly degenerate in energy, see
Table~\ref{tab:Grav}.  The one with lowest energy is shown in
Fig.~\ref{fig:9and10}.  It has two four-fold and eight five-fold
corners.  It has a special symmetry which is a simultaneous reflection
in the horizontal plane and a rotation by $45^{\circ}$.  It also has
four vertical symmetry planes.

The second stable configuration with ten particles, with slightly
higher energy, is just the nine-particle configuration shown in
Fig.~\ref{fig:9and10} with a tenth particle in the centre.  This is
actually the first time we encounter a minimum energy configuration
with a central particle.  It is also the first time we encounter two
nearly degenerate minimum energy configurations.

{\bf 11 particles.}  The configuration with eleven particles is simply
the one with ten particles shown in Fig.~\ref{fig:9and10} with the
eleventh particle in the centre.

\begin{figure}
\begin{center}
\fbox {\includegraphics[width=7.48cm]{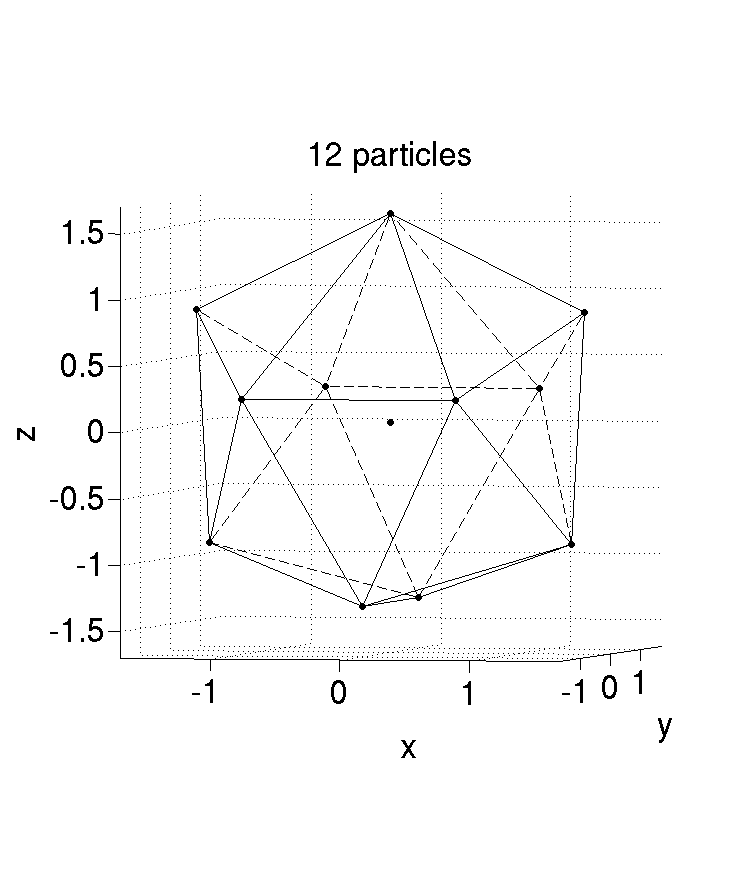}}
\fbox {\includegraphics[width=7.48cm]{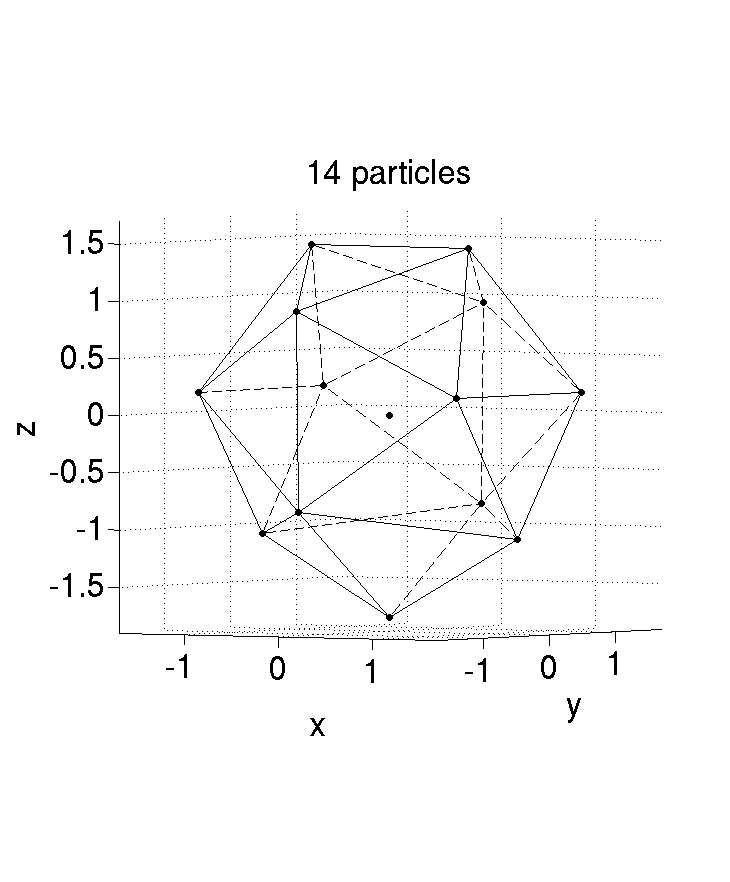}}
\caption{\label{fig:12and14} Minimum energy configurations.}
\end{center}
\end{figure}

{\bf 12 particles.}  The minimum energy configuration with twelve
particles is shown in Fig.~\ref{fig:12and14}, It has one central
particle at $x=y=0$ and $z=0.034$, slightly above the horizontal plane
because this is not a symmetry plane.  The only symmetries are
reflection symmetries about the $xz$- and $yz$-planes.  The top corner
is six-fold.  Two corners adjacent to this are four-fold, and the
remaining eight corners are five-fold.

{\bf 13 particles.}  The minimum energy configuration with thirteen
particles is one of the few that have maximal symmetry.  It is a
regular icosahedron with the thirteenth particle in the centre.

{\bf 14 particles.}  In Fig.~\ref{fig:12and14} we show also the
configuration with fourteen particles.  The bottom corner is
four-fold, and the remaining twelve corners are five-fold.  Again the
central particle is not exactly at the centre, because the horizontal
plane is not a symmetry plane.  The only symmetry is a $180^{\circ}$ rotation
about the $z$-axis.

\begin{figure}
\begin{center}
\fbox {\includegraphics[width=7.48cm]{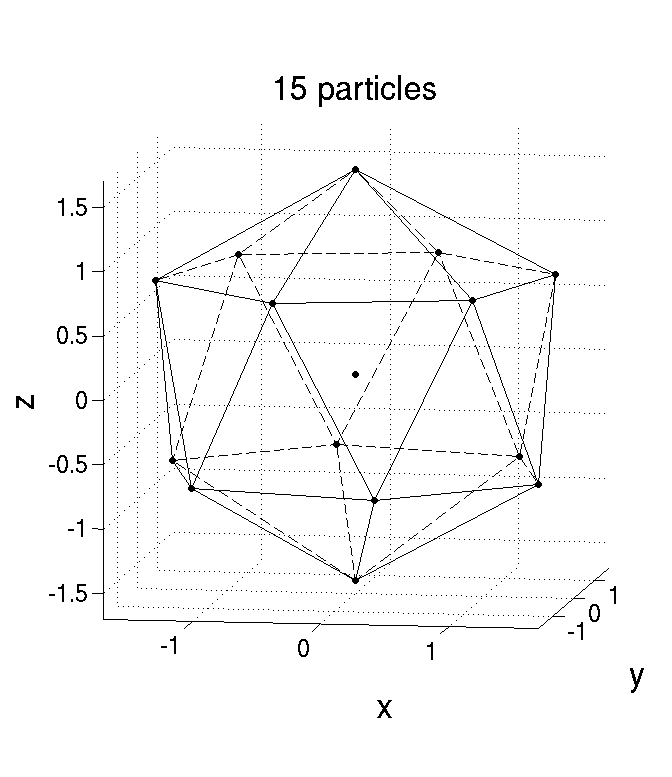}}
\caption{\label{fig:15} Minimum energy configuration with a particle
  exactly at the centre.}
\end{center}
\end{figure}

{\bf 15 particles.}  In Fig.~\ref{fig:15} we show the configuration of
fifteen particles.  The central particle is exactly at the centre.
The top and bottom corners are six-fold, and the remaining twelve
corners are five-fold.  A simultaneous reflection about the horizontal
plane and a rotation by $30^{\circ}$ about the $z$-axis is a
symmetry. In addition there are six vertical symmetry planes.

\begin{figure}
\begin{center}
\fbox {\includegraphics[width=7.48cm]{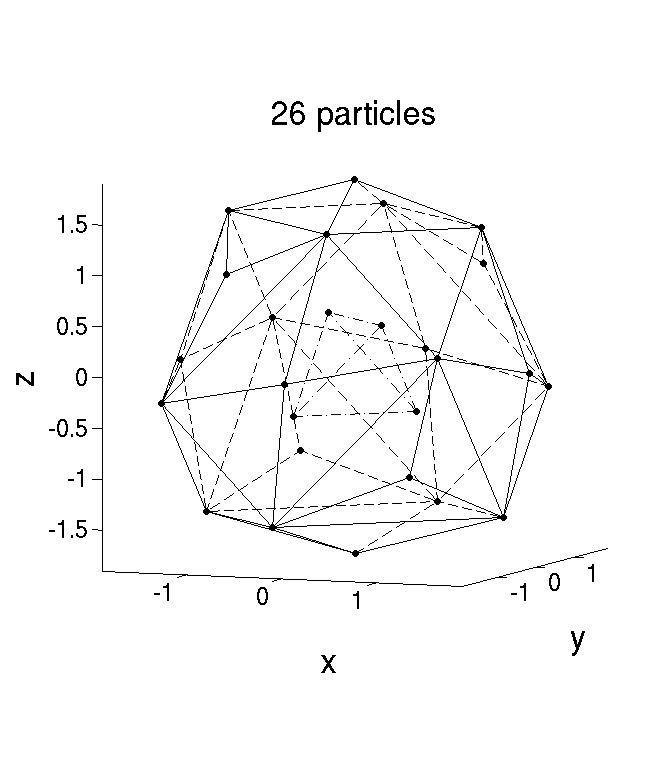}}
\fbox {\includegraphics[width=7.48cm]{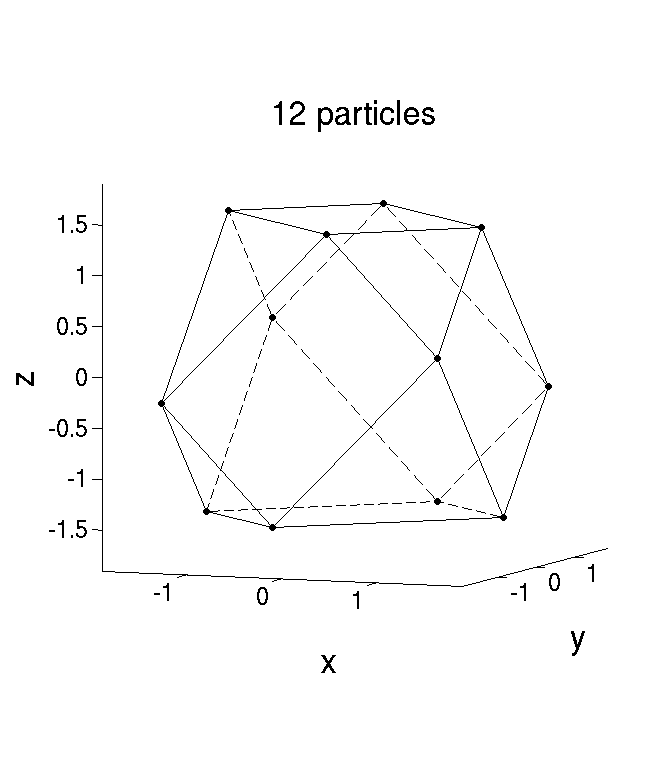}}
\\
\fbox {\includegraphics[width=7.48cm]{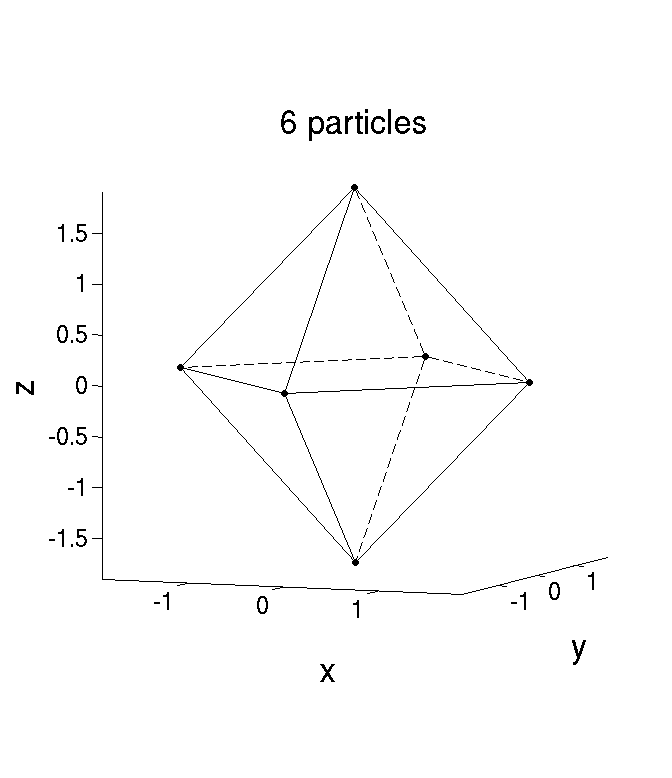}}
\fbox {\includegraphics[width=7.48cm]{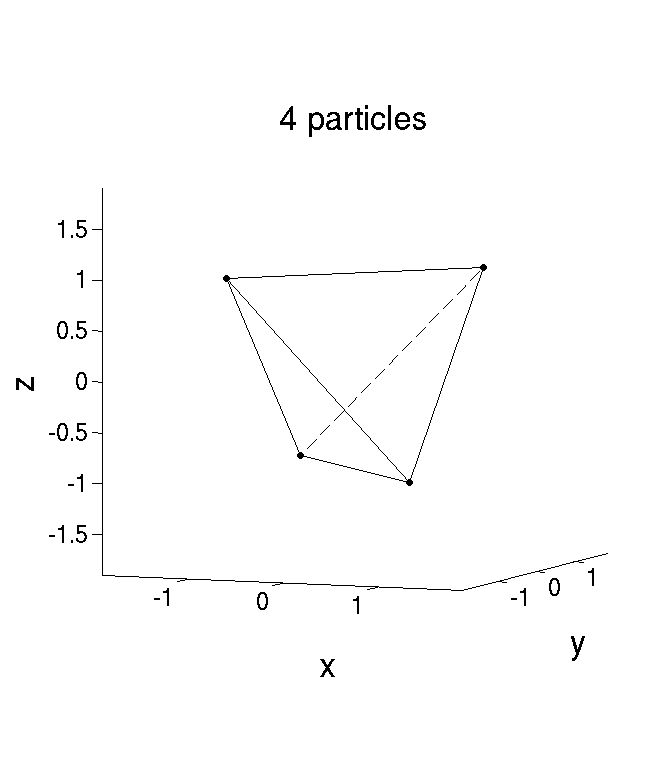}}
\caption{\label{fig:26} Minimum energy configuration with 26
  particles, consisting of an inner tetrahedron and and an outer shell
  made up of three simple polyhedra.}
\end{center}
\end{figure}

{\bf 26 particles.}  In Fig.~\ref{fig:26} we show the ``magic''
configuration of 26 particles.  This has an exceptionally low energy
because of its high symmetry, which is not obvious from the complete
figure, although the projections onto the $xy$-, $xz$- and $yz$-planes
are equal and highly symmetric.  There is a central regular
tetrahedron, and the symmetry group of the whole configuration
consists of all the 24 permutations of the corners of this
tetrahedron.  The tetrahedron is surrounded by a shell made up of
three different polyhedra with the same tetrahedral symmetries, shown
here in the same figure.  One contains 12 particles and can be
understood as a cube with all six corners cut off.  The other two are
a regular octahedron and a regular tetrahedron

\begin{figure}
\begin{center}
\fbox {\includegraphics[width=7.48cm]{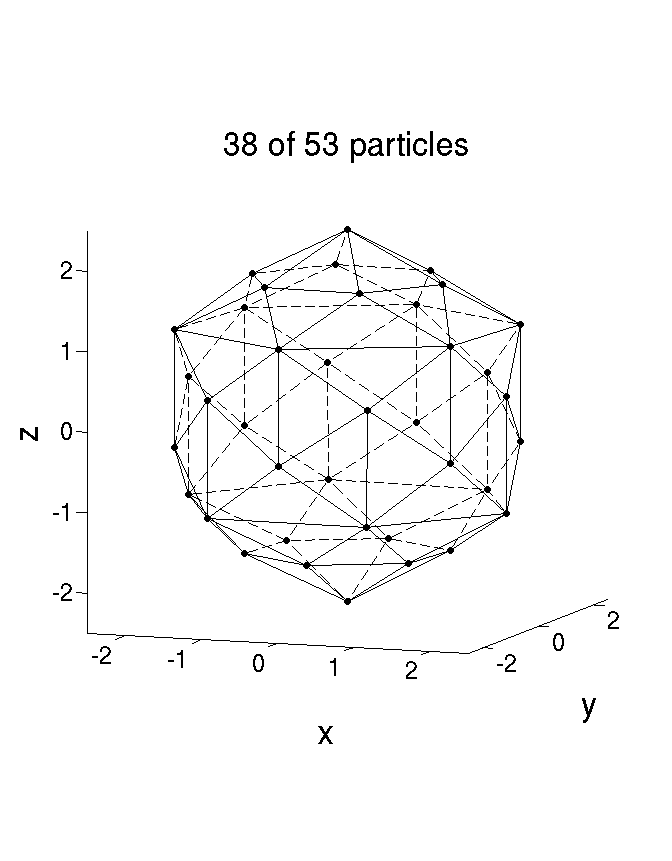}}
\fbox {\includegraphics[width=7.48cm]{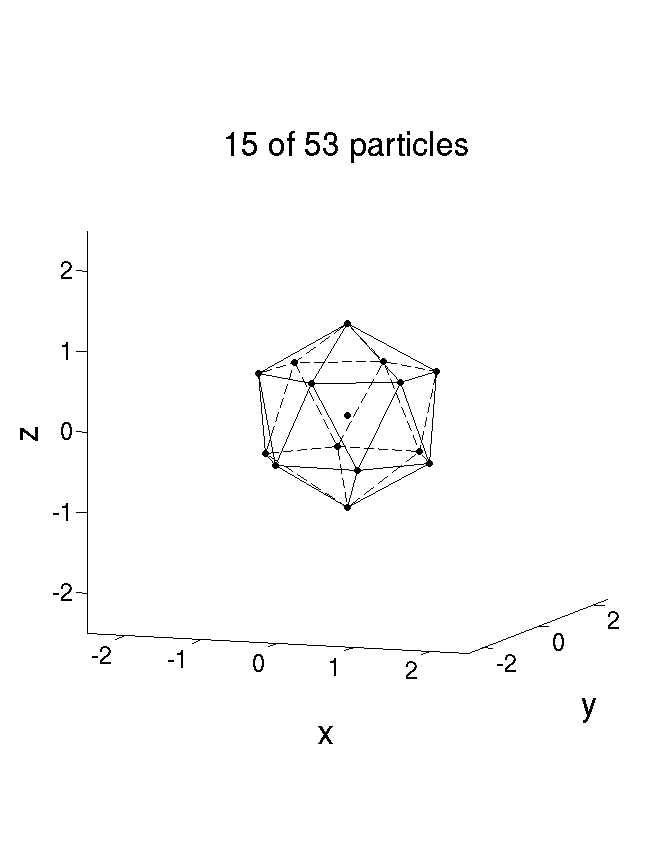}}
\caption{\label{fig:53} Minimum energy configuration with 53
  particles.  There is one central particle, an inner shell of 14
  particles (right), and an outer shell of 38 particles (left),}
\end{center}
\end{figure}

{\bf 53 particles.}  In Fig.~\ref{fig:53} we show the configuration of
53 particles, another ``magic'' case.  There is one central particle
exactly at the centre, and two concentric approximately spherical
shells, plotted separately in the figure.  The inner shell contains 14
particles and has a root mean square radius of $1.211\,011\,415\,841$.
The outer shell contains 38 particles and has a radius of
$2.237\,680\,879\,671$.  The inner shell plus the central particle is
like the configuration of 15 particles plotted in Fig.~\ref{fig:15},
compressed by a factor of $0.71$.  In fact, the root mean square
radius of the shell in Fig.~\ref{fig:15} is $1.707\,609\,454\,071$.

A complete description of the configuration of 53 particles is given
in Table~\ref{tab:53points}.  Five particles lie along the $z$-axis,
while 48 particles form regular hexagons centered on the $z$-axis in
eight planes perpendicular to the $z$-axis.  The hexagons in adjacent
planes are rotated $30^{\circ}$ relative to each other.  In the table,
$r_2$ is the distance of the points from the $z$ axis (the radius of
the hexagons), and $r_3$ is the distance from the origin.  The basic
symmetry transformation is a simultaneous rotation by $30^{\circ}$
about the $z$-axis and a reflection about the $xy$-plane.  In
addition, there are six vertical symmetry planes.

\begin{table}[h]
\begin{center}
\begin{tabular}{|r|r|r|}
\hline
\multicolumn{1}{|c}{$z$}  &\multicolumn{1}{|c}{$r_2$} & \multicolumn{1}{|c|}{$r_3$}\\
\hline
 $    0.0\phantom{00\,000\,000\,000}$ & $0.0\phantom{00\,000\,000\,000}$ &
     $0.0\phantom{00\,000\,000\,000}$\\
 $\pm 1.140\,738\,013\,178$ & $0.0\phantom{00\,000\,000\,000}$ & $1.140\,738\,013\,178$\\
 $\pm 2.304\,942\,769\,817$ & $0.0\phantom{00\,000\,000\,000}$ & $2.304\,942\,769\,817$\\
\hline
 $\pm 0.360\,452\,820\,441$ & $2.138\,438\,168\,526$ & $2.168\,604\,167\,747$\\
 $\pm 0.535\,171\,243\,082$ & $1.098\,947\,054\,672$ & $1.222\,330\,924\,257$\\
 $\pm 1.098\,246\,666\,131$ & $2.135\,949\,636\,872$ & $2.401\,754\,898\,178$\\
 $\pm 1.697\,413\,737\,050$ & $1.272\,100\,190\,070$ & $2.121\,191\,242\,746$\\
\hline
\end{tabular}
\caption{\label{tab:53points} The minimum energy configuration for
  $53$ particles without rotation.  $r_2$ and $r_3$ are defined in the
  text.}
\end{center}
\end{table}

\section{Rotating systems of finite numbers of point particles}

\label{sec:5}

So far we have studied nonrotating minimum energy configurations,
arguing that they can represent polytropes of index $n=3/2$, as
described by the Lane--Emden equation.  In particular, we have shown
that already with 120 point particles they reproduce very well the
polytropic density profile.
We will now show how they become deformed when set into rotation.

In contrast to a liquid of constant density, a polytrope is
compressible.  It has a sharply defined surface, like the liquid, but
its density goes to zero at the surface.
Jeans~\cite{JeansI,JeansII,JeansIII} concluded that the
compressibility implies that the transition from a Maclaurin
ellipsoid, having a circular shape in the plane perpendicular to the
rotation axis, to a Jacobi ellipsoid, having three different principal
axes, can take place with a rotating polytrope only when the
polytropic index $n$ is smaller than a critical value
$n_c\approx 0.83$.  This value was confirmed by James~\cite{James},
who found $n_c\approx 0.808$.  See Appendix~\ref{app:Jeans} for a
summary of the reasoning of Jeans.

It means that we can not expect to see the Maclaurin to Jacobi
transition in our simulations with $n=1.5$.  What happens instead,
before the bifurcation point is reached, is that the rotating
polytrope becomes unstable by shedding particles at the equator, where
the centrifugal force exactly balances the gravitational attraction.
This balance implies that the equator becomes a sharp edge, as
described by Jeans.  In our simulations, as presented here, we see
clearly this sharp edge appearing when the angular momentum reaches
its maximal value.

\subsection{The maximal angular momentum}

\begin{figure}[h]
\begin{center}
\includegraphics[width=9cm]{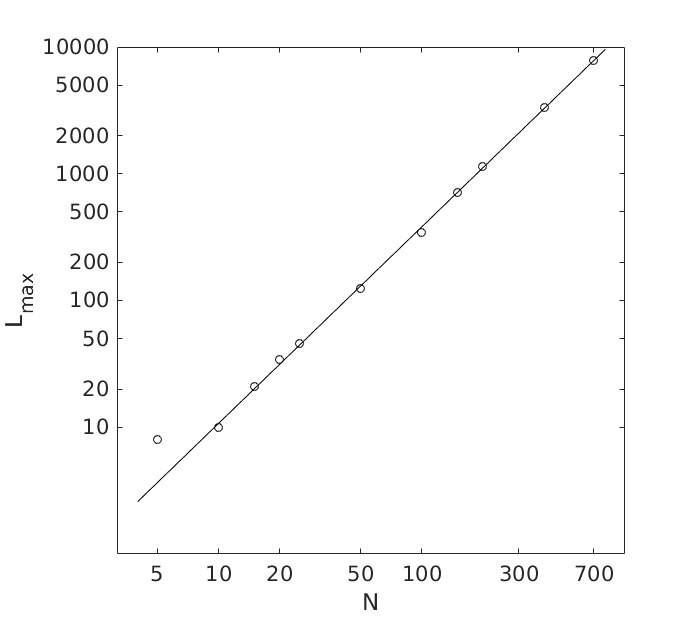}
\caption{\label{fig:Lmax} The numerical values of $L_{\textrm{max}}$
  given in Table~\protect{\ref{tab:Lmax}}.  The straight line is the
  power law given in Eq.~(\protect{\ref{eq:Lmax}}).}
\end{center}
\end{figure}

We have tried to determine with good precision, for some values of
$N$, the maximal value $L_{\textrm{max}}$ that the angular momentum
$L$ can take before the system becomes unstable by losing particles.
The results are summarized in Table~\ref{tab:Lmax} and
Fig,~\ref{fig:Lmax}.  The figure shows that the data follow very
closely the power law
\be
\label{eq:Lmax}
L_{\textrm{max}} = aN^b\;,
\ee
with $a=0.30157$ and $b=1.5510$.

\begin{table}[H]
\begin{center}
\begin{tabular}{ccrc}
$N$ & $L_\textrm{max}$ &  $W_\textrm{max}$\\
\hline
\phantom{00}5 &                         $\phantom{000}8$&    $0.1876$&\\
\phantom{0}10 &                         $\phantom{00}10$&    $0.0733$&\\
\phantom{0}15 &                         $\phantom{00}21$&    $0.0804$&\\
\phantom{0}20 &                         $\phantom{00}34$&    $0.0827$&\\
\phantom{0}25 &                         $\phantom{00}46$&    $0.0757$&\\
\phantom{0}50 &                         $\phantom{0}130$&    $0.0647$&\\
100 &                                   $\phantom{0}345$&    $0.0595$&\\
150 &                                   $\phantom{0}715$&    $0.0681$&\\
200 &                                   $1140$&              $0.0701$&\\
400 &                                   $3348$&              $0.0711$&\\
700 &                                   $7895$&              $0.0713$&
\end{tabular}
\caption{\label{tab:Lmax} The maximal angular momentum $L_{\textrm{max}}$ for 
different numbers of particles, and the corresponding $W$ values.
Note the drop in $W_{\textrm{max}}$ from $N=5$ to $N=10$. }
\end{center}
\end{table}

We can understand the power $b\approx 1.5$ in the following way.  In
the virial theorem, Eq.~(\ref{eq:virth}), we have for large $N$ that
the first two terms are proportional to the number of particle pairs,
$N(N-1)/2$,
\be
U_{\textrm{rep}}=AN^2\;,\qquad
V_{\textrm{grav}}=-BN^2\;,
\ee
with positive coefficients $A$ and $B$ that do not vary much with $N$.
In the rotational energy $E_{\textrm{rot}}=L^2/(2I)$, the moment of
inertia is proportional to the number of particles,
\be
I=CN\;,
\ee
with a coefficient $C$ that is again nearly independent of $N$.  Thus
the virial theorem implies that
\be
L=\sqrt{C(B-A)N^3}\;.
\ee
This argument would imply that $b\to 1.5$ in the limit $N\to\infty$.

\subsection{The ratio between rotational and gravitational energy }

The table also gives the maximal value $W_{\textrm{max}}$ of the ratio
between the rotational energy and the gravitational energy,
$W=E_{\textrm{rot}}/|V_{\textrm{grav}}|$.  This maximal value is known
to be smaller for a polytrope than for a liquid of constant
density~\cite{Shap}.  It is an interesting result we obtain that the
value of $W_{\textrm{max}}$ seems to be roughly independent of the
number of particles.

The virial theorem gives Eq.~(\ref{eq:16}),
\be
\label{eq:16b}
W=\frac{E_{\rm rot}}{|V_{\rm grav}|}
=\frac{1}{2}-\frac{U_{\rm rep}}{|V_{\rm grav}|}
\leq\frac{1}{2}\;.
\ee
This upper limit of $1/2$ is much larger than the values
$W_{\textrm{max}}\approx 0.07$ that we find in our simulations.  The
exceptional value $W_{\textrm{max}}=0.1876$ listed in the table for
five particles is not representative, because it occurs when the
particles line up on a straight line, and
${U_{\rm rep}}/{|V_{\rm grav}|}\approx 0.31$, see~\cite{HM}.  The low
value $W_{\textrm{max}}\approx 0.07$ corresponds to a much higher
value of the repulsive potential,
${U_{\rm rep}}/{|V_{\rm grav}|}\approx 0.43$.

Note that the upper limit of $1/2$ in Eq.~(\ref{eq:16b}) is the same
as the upper limit for a Maclaurin ellipsoid in the case of an
incompressible fluid~\cite{Shap}.  We could in principle simulate a
nearly incompressible fluid by inserting a very large value for the
power $k=3(\gamma-1)$ in Eq.~(\ref{eq:Urep}).  This would imply a
generalized virial theorem, and the following generalization of
Eq.~(\ref{eq:16b}),
\be
W=\frac{E_{\rm rot}}{|V_{\rm grav}|}
=\frac{1}{2}-\frac{kU_{\rm rep}}{2\,|V_{\rm grav}|}
\leq\frac{1}{2}\;.
\ee
We may argue that it would also lead to a small limiting value for
${kU_{\rm rep}}/{|V_{\rm grav}|}$, so that $W$ would approach the
limit of $1/2$.

\section{Example: 400 particles}
\label{sec:6}

We have generated more than 1600 configurations consisting of 400
point particles with different values of the angular momentum.  Most
of them are generated by following stable branches, increasing or
decreasing the angular momentum $L$ in small steps.  In this section
we present some results obtained from studying these data.  For
comparison we also present some results for 700 particles, and also
some cases of fewer than 400 particles.

The general picture is that with an increasing number of particles the
systems we study resemble more and more a continuous polytrope.  We
confirm the Jeans effect as defined in the introduction, that
configurations become unstable at the highest values of $L$ by
shedding single particles at the equator, which then becomes a sharp
edge.  We also see strong evidence for the result obtained by Jeans
that a polytrope with polytropic index $n=1.5$ remains rotationally
symmetric in the rotation plane until it becomes unstable at
$L=L_{\textrm{max}}$.

\subsection{Stability} 

\begin{figure}
  \begin{center}
\fbox {\includegraphics[width=9cm]{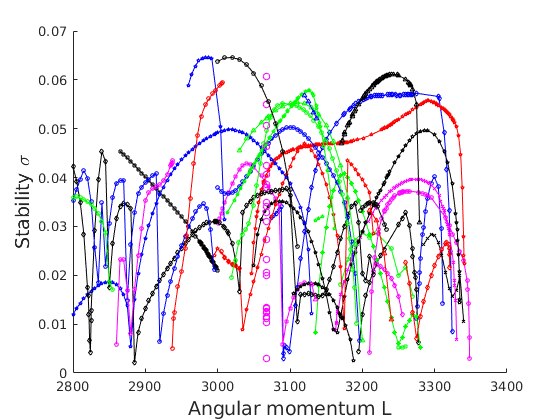}}    
\caption{\label{fig:stab} Evolution of the stability parameter
  $\sigma$ as a function of the angular momentum $L$, for around 40
  stable branches of configurations of 400 particles.  The isolated
  circles at $L=3067$, belonging to other branches that we have not
  followed, indicate that the present sample of branches is very far
  from complete.}
\end{center}
\end{figure}

In paper I we studied systems of very few particles, and gave detailed
descriptions of how different branches of stable configurations evolve
with increasing angular momentum $L$.  Obviously, this we can not do
here in the same detail, because the number of branches is extremely
large.  Nevertheless, we can pick at random one branch at a time and
follow its evolution, changing $L$ in small steps.  In
Fig.~\ref{fig:stab} we plot the stability parameter $\sigma$ for 400
particles, as a function of $L$, for some 40 branches above $L=2800$.
Every branch is seen to be stable only in a rather small $L$ interval.
We find no stable branches above $L_{\textrm{max}}=3348$.

As an indication of to what degree our random sample of branches is
complete, we have searched for, and found, other stable configurations
at $L=3067$.  These are plotted in the figure as (purple) circles,
They indicate that continued searches would reveal a very large number
of branches, also in seemingly empty areas in the plot.

\subsection{ Asymmetry } 

\begin{figure}
  \begin{center}
\fbox {\includegraphics[width=9cm]{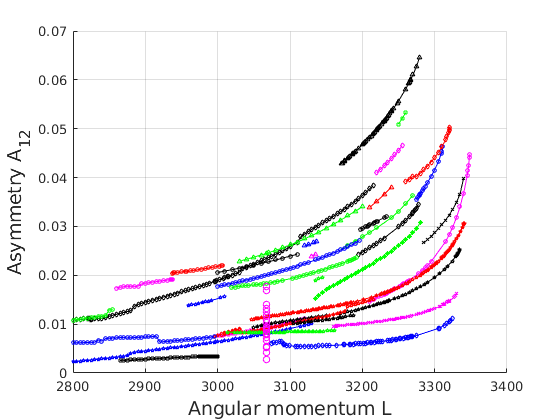}}
\caption{\label{fig:A12400} This plot for 400 particles shows how the
  asymmetry parameter $A_{12}$ increases with the angular momentum
  $L$.  The sample of branches, and the isolated points at $L=3067$,
  are the same as in Fig.~\protect{\ref{fig:stab}}.}  
\end{center}
\end{figure}

In Section~\ref{subsec:Asymmetry} we defined different asymmetry
parameters.  In Fig.~\ref{fig:A12400} we see how the asymmetry
$A_{12}$, in the rotation plane (the $xy$-plane), increases with $L$.
The main result to be noted is that it is small.  We believe that the
reason for the sharp increase in $A_{12}$ when $L$ comes close to
$L_{\textrm{max}}$, is that the number of particles is finite, and
single particles at the equator become more and more loosely bound.
This is simply what we call the Jeans effect.

\begin{figure}
\begin{center}
\fbox {\includegraphics[width=7.48cm]{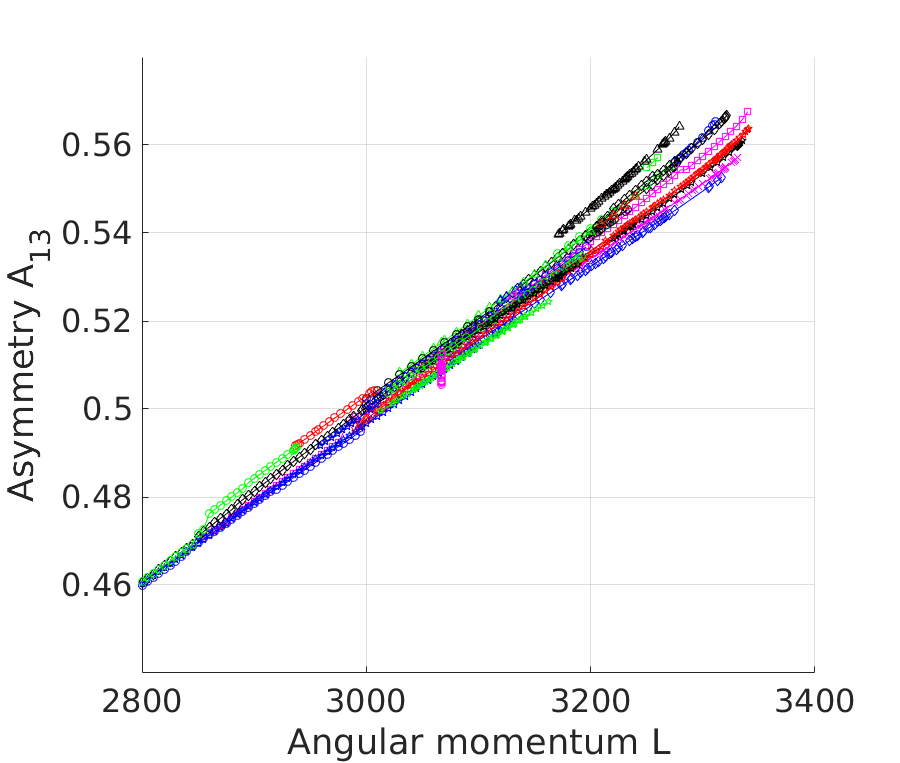}}
\fbox {\includegraphics[width=7.48cm]{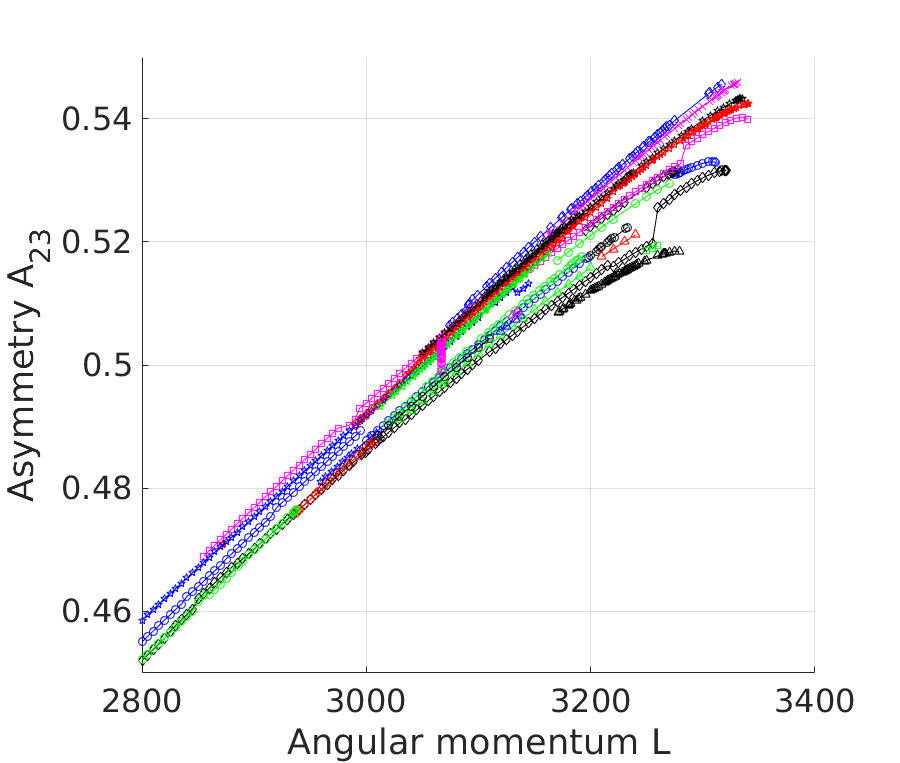}}
\caption{\label{fig:assyxz} These plots for 400 particles show how the
  asymmetry parameters $A_{13}$, in the left panel, and $A_{23}$, in
  the right panel, evolve as functions of the angular momentum $L$ .}
\end{center}
\end{figure}

In the left panel of Fig.~\ref{fig:assyxz} we see how the asymmetry
$A_{13}$, in the $xz$-plane, changes as a function of $L$.  The right
panel shows the asymmetry $A_{23}$, in the $yz$-plane.  They are both
much larger than the $xy$ asymmetry $A_{12}$.  The reason is of course
that they measure the rotational flattening, which is large.  The
smallness of $A_{12}$ in comparison indicates that it may vanish in
the continuum limit $N\to\infty$, as predicted by Jeans.  For a given
value of $L$ approaching $L_{\textrm{max}}$, all three asymmetries, in
particular $A_{12}$, show substantial variations.

\begin{figure}
  \begin{center}
\fbox {\includegraphics[width=9cm]{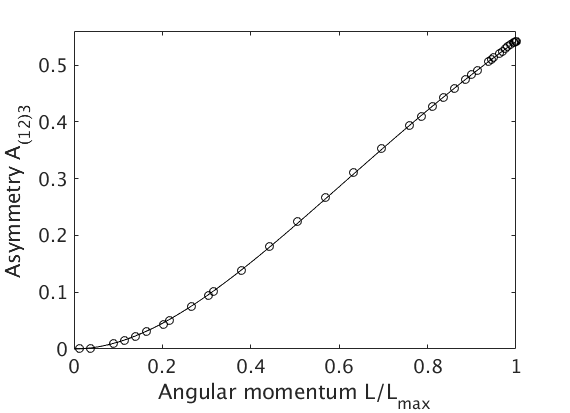}}
\caption{\label{fig:asymstab4} The asymmetry parameter $A_{(12)3}$ as
  a function of the scaled angular momentum $L/L_{\textrm{max}}$ for
  700 particles.  Here $L_{\textrm{max}}=7895$.  The curve is a quartic
  polynomial fitted to the data points, see
  Eq.~(\protect{\ref{eq:cubicfit400}}).}
\end{center}
\end{figure}

\begin{figure}
  \begin{center}
\fbox {\includegraphics[width=7.48cm]{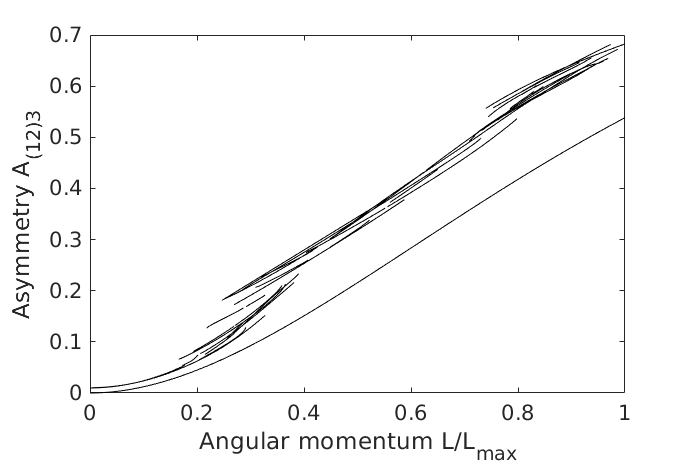}}
\fbox {\includegraphics[width=7.48cm]{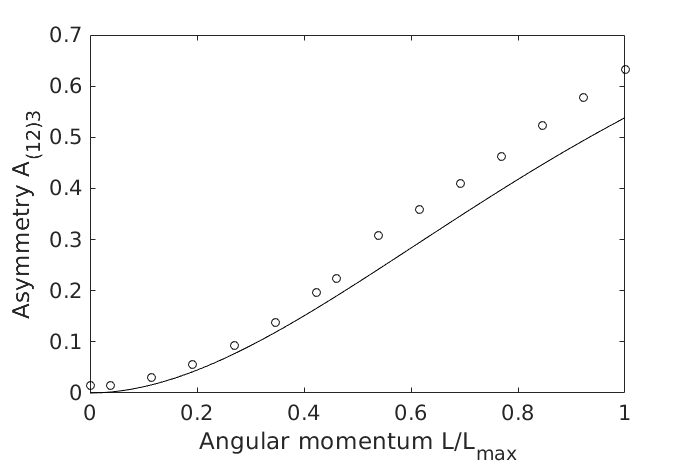}}    
\caption{\label{fig:asymstab} The asymmetry parameter $A_{(12)3}$ as a
  function of the angular momentum $L/L_{\textrm{max}}$ for 25
  particles, with $L_{\textrm{max}}=44.6$, and for 50 particles,
  with $L_{\textrm{max}}=130$.  The curve in both plots is the
  quartic polynomial fitted to the data for 700 particles.}
\end{center}
\end{figure}

\begin{figure}
\begin{center}
\fbox {\includegraphics[width=7.48cm]{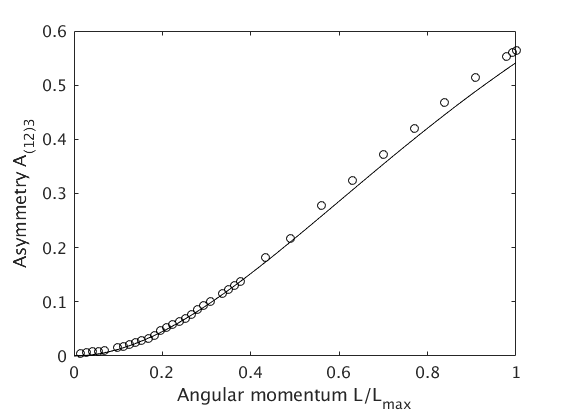}}
\fbox {\includegraphics[width=7.48cm]{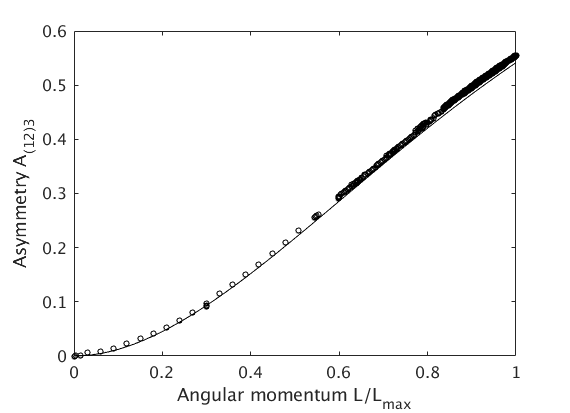}}
\caption{\label{fig:asymstab2} The asymmetry parameter $A_{(12)3}$ as a
  function of the angular momentum $L/L_{\textrm{max}}$ for 150 particles,
  with $L_{\textrm{max}}=715$,
  and for 400 particles, with $L_{\textrm{max}}=3348$.  The curve  is the
  fit for 700 particles.}
\end{center}
\end{figure}

It follows from Eq.~(\ref{eq:asymrelations}) that $A_{13}=A_{23}$ if
$A_{12}=0$, and that $A_{13}>A_{23}$ if $A_{12}>0$.  Since we have
$A_{12}>0$ in our data, and expect that $A_{12}=0$ in the continuum
limit, we conclude that the proper asymmetry to study in the limit is
neither $A_{13}$ nor $A_{23}$, but rather $A_{(12)3}$ as defined in
Eq.~(\ref{eq:A123def}).  This is plotted in Fig.~\ref{fig:asymstab4}
for 700 particles, together with a fitted curve which is the following
quartic polynomial,
\be
\label{eq:cubicfit400}
A_{\textrm{fit}}
=c_2\left(\frac{L}{L_{\textrm{max}}}\right)^{\!2}
+c_3\left(\frac{L}{L_{\textrm{max}}}\right)^{\!3}
+c_4\left(\frac{L}{L_{\textrm{max}}}\right)^{\!4}\;,
\ee
with $L_{\textrm{max}}=7895$ and with
\be
c_2 =  1.33288\;,\qquad
c_3 = -1.05954\;,\qquad
c_4 =  0.26853\;.
\ee
We have assumed in the fit that the asymmetry grows quadratically with
$L$ for small $L$.  The fit is remarkably good with as few as three
parameters.

The Figures~\ref{fig:asymstab} and~\ref{fig:asymstab2} show the same
asymmetry $A_{(12)3}$ as a function of $L/L_{\textrm{max}}$ for 25,
50, 150, and 400 particles, compared to the curve fitted for 700
particles.  It is worth noting that for 400 particles, where we have
1628 data points for $A_{(12)3}$, all these points fall very nearly on
one single curve, with fluctuations presumably because there are a
finite number of particles.  We see in Fig.~\ref{fig:assyxz} that
$A_{13}$ and $A_{23}$ show larger variations for a fixed value of $L$
(but note the factor of nearly ten between the scales of the plots).

If there does exist a continuum limit when $N\to\infty$ we would
expect the asymmetry as a function of $L/L_{\textrm{max}}$ to become
independent of $N$ in the limit.  These plots show clearly a finite
size effect, that the asymmetry decreases when the number of particles
increases.  We believe that the continuum limit exists, but we have
not quite reached it yet.

\subsection{The energy as a function of $L$} 

\begin{figure}
\begin{center}
\fbox {\includegraphics[width=7.48cm]{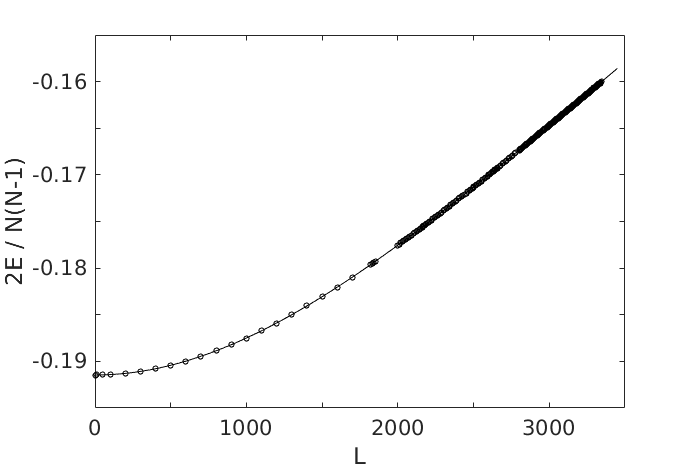}}
\fbox {\includegraphics[width=7.48cm]{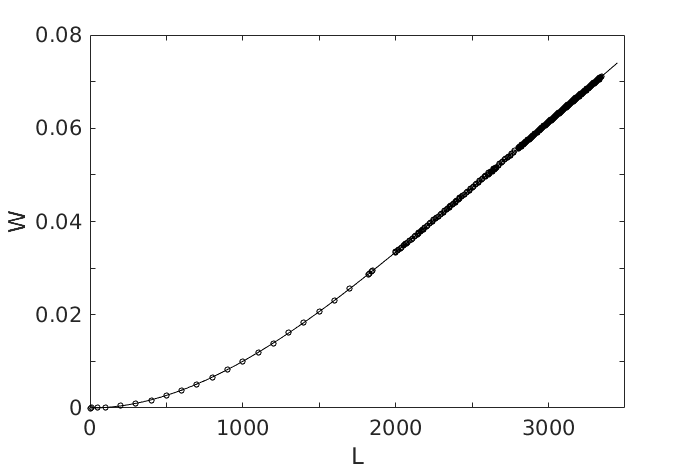}}
\caption{\label{fig:12and14c} The left panel shows the average energy per
  particle pair as a function of the angular momentum $L$, for
  $N=400$.  The curve is the fitted polynomial given in
  Eq.~(\protect{\ref{eq:quinticfit400}}).  The right panel shows the
  ratio $W$ as a function of $L$.  Again the curve is a polynomial
  fit, Eq.~(\protect{\ref{eq:quinticfit400a}}).}
\end{center}
\end{figure}

The left panel in Fig.~\ref{fig:12and14c} shows how the average energy per
particle pair, $\langle E_2\rangle=2E/N(N-1)$, increases with the
angular momentum $L$ for 400 particles.  The following polynomial of
degree five gives an excellent fit,
\be
\label{eq:quinticfit400}
E_{\textrm{fit}}
=d_0
+d_2\left(\frac{L}{L_{\textrm{max}}}\right)^{\!2}
+d_3\left(\frac{L}{L_{\textrm{max}}}\right)^{\!3}
+d_4\left(\frac{L}{L_{\textrm{max}}}\right)^{\!4}
+d_5\left(\frac{L}{L_{\textrm{max}}}\right)^{\!5}\;,
\ee
with $L_{\textrm{max}}=3348$ and with
\be
d_0 = -0.19146\;,\quad
d_2 =  0.04881\;,\quad
d_3 = -0.01170\;,\quad
d_4 = -0.01118\;,\quad
d_5 =  0.00556\;.
\ee
The fitted value of $d_0$, the energy at $L=0$, is the same as the
value at $N=400$, $L=0$, given in Table~\ref{tab:Grav}.

The right panel in Fig.~\ref{fig:12and14c} shows how the ratio of
rotational and gravitational energy,
$W=E_{\textrm{rot}}/|V_{\textrm{grav}}|$, increases with $L$.  We fit
it by the following polynomial,
\be
\label{eq:quinticfit400a}
W_{\textrm{fit}}
=e_2\left(\frac{L}{L_{\textrm{max}}}\right)^{\!2}
+e_3\left(\frac{L}{L_{\textrm{max}}}\right)^{\!3}
+e_4\left(\frac{L}{L_{\textrm{max}}}\right)^{\!4}
+e_5\left(\frac{L}{L_{\textrm{max}}}\right)^{\!5}\;,
\ee
with
\be
e_2 =  0.12960\;,\quad
e_3 = -0.05269\;,\quad
e_4 = -0.02195\;,\quad
e_5 =  0.01618\;.
\ee

\subsection{The Jeans effect}

\begin{figure}
\begin{center}
\fbox {\includegraphics[width=7.2cm]{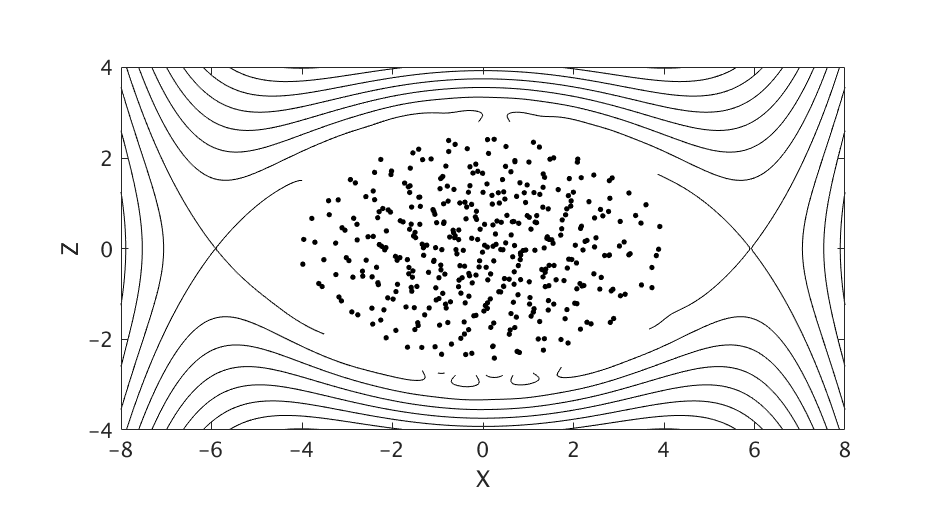}}
\fbox {\includegraphics[width=5.49cm]{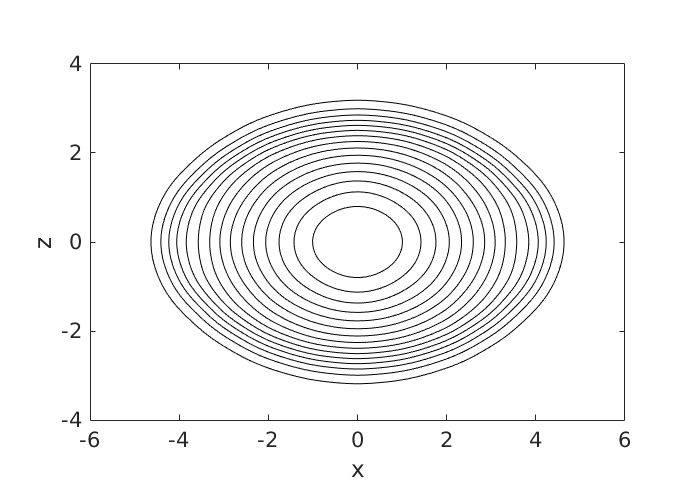}}
\fbox {\includegraphics[width=7.2cm]{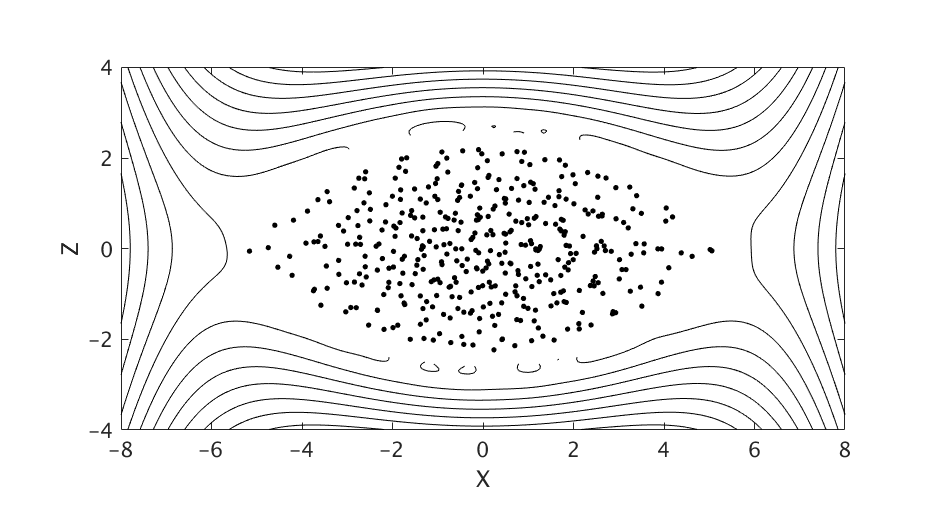}}
\fbox {\includegraphics[width=5.49cm]{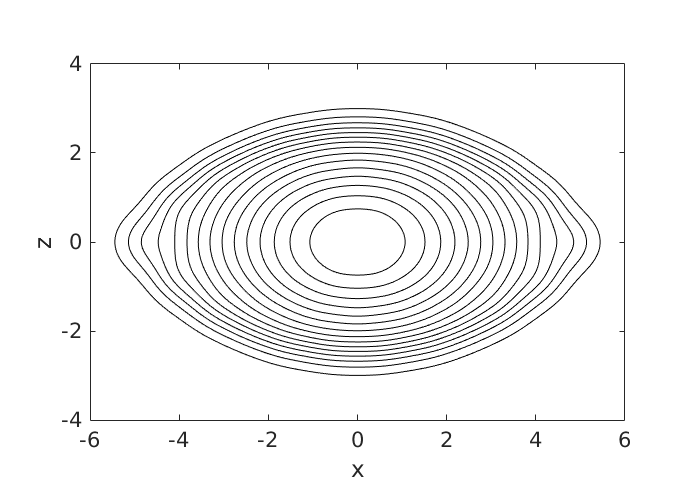}}    
\caption{\label{fig:12and14d} The left panels show the $xz$
  projections of two different configurations with 400 particles, at
  angular momenta $L=2500$ (top) and $L=L_{\textrm{max}}=3348$
  (bottom).  Also plotted are equipotential lines in the $xz$-plane.
  The right panels show the same configurations represented as contour
  curves of the densities in the $xz$-plane.  The sharp edge at the
  equator at the maximum angular momentum is clearly seen.}
\end{center}
\end{figure}

\begin{figure}
\begin{center}
\fbox {\includegraphics[width=7.2cm]{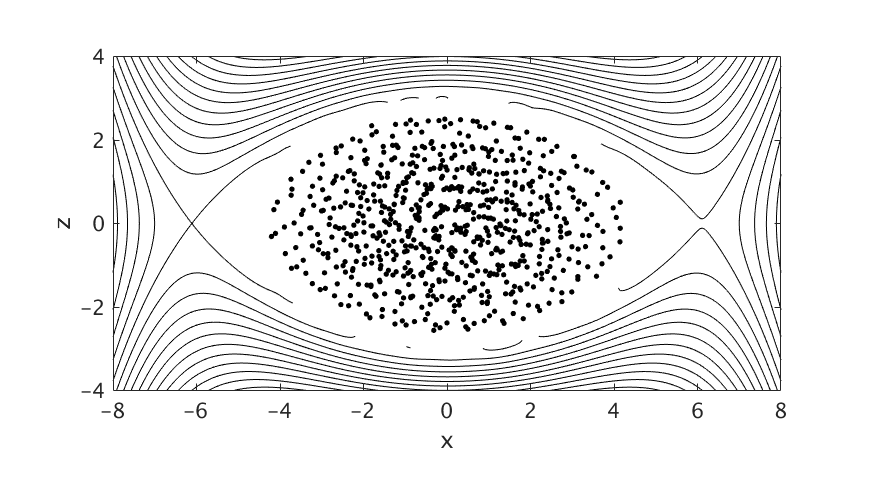}}
\fbox {\includegraphics[width=5.49cm]{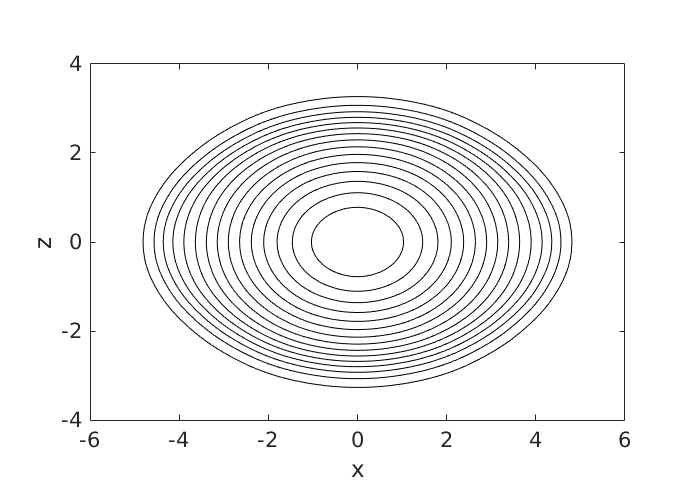}}
\fbox {\includegraphics[width=7.2cm]{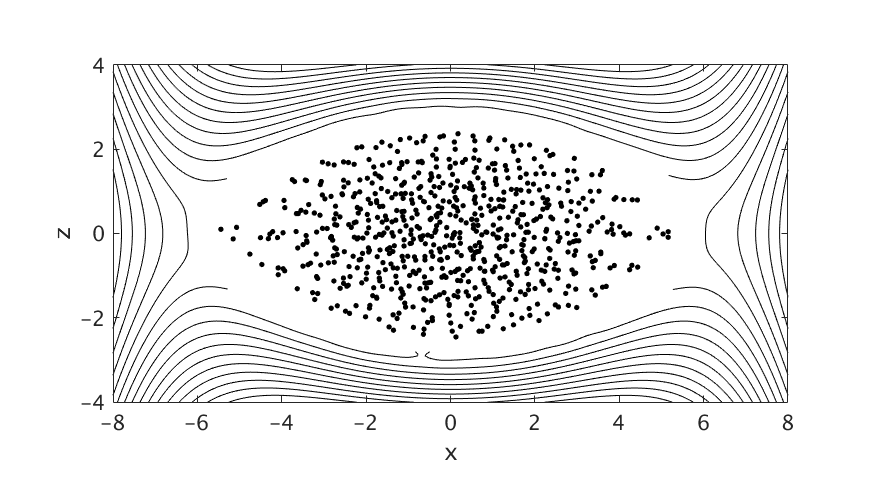}}
\fbox {\includegraphics[width=5.49cm]{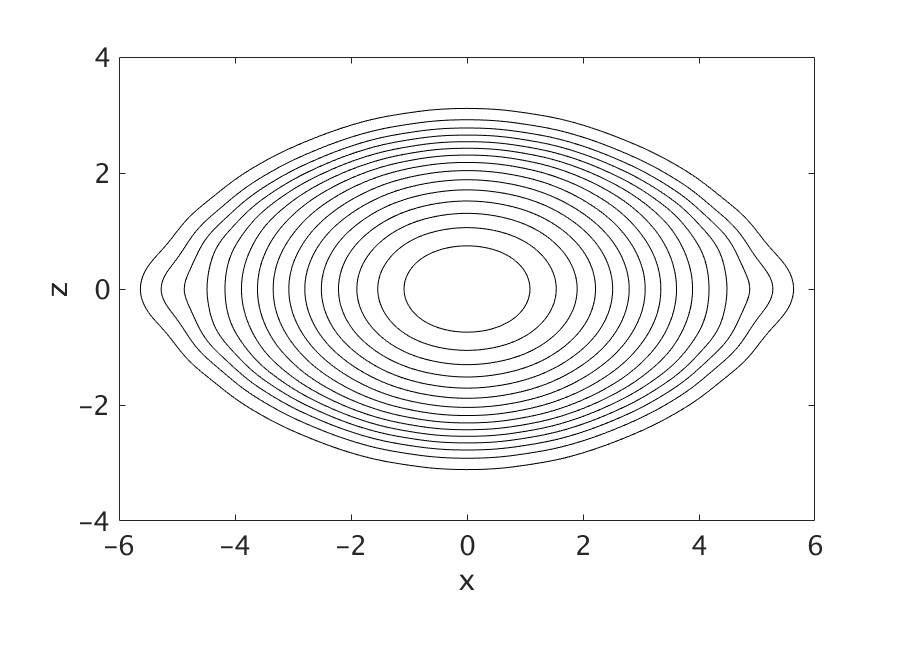}}
\caption{\label{fig:12and14dd} The left panels show the $xz$
  projections of two configurations of 700 particles, at $L=6000$
  (top) and at $L=L_{\textrm{max}}=7895$ (bottom).  The right panels
  show the same configurations represented as contour curves of
  the densities.  Again the sharp edge at the equator at the maximum
  angular momentum is clearly seen.}
\end{center}
\end{figure}

\begin{figure}
\begin{center}
\fbox {\includegraphics[width=6.48cm]{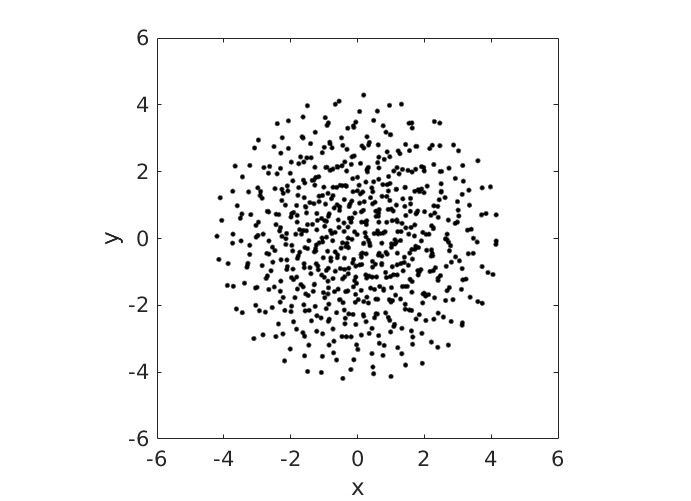}}
\fbox {\includegraphics[width=6.48cm]{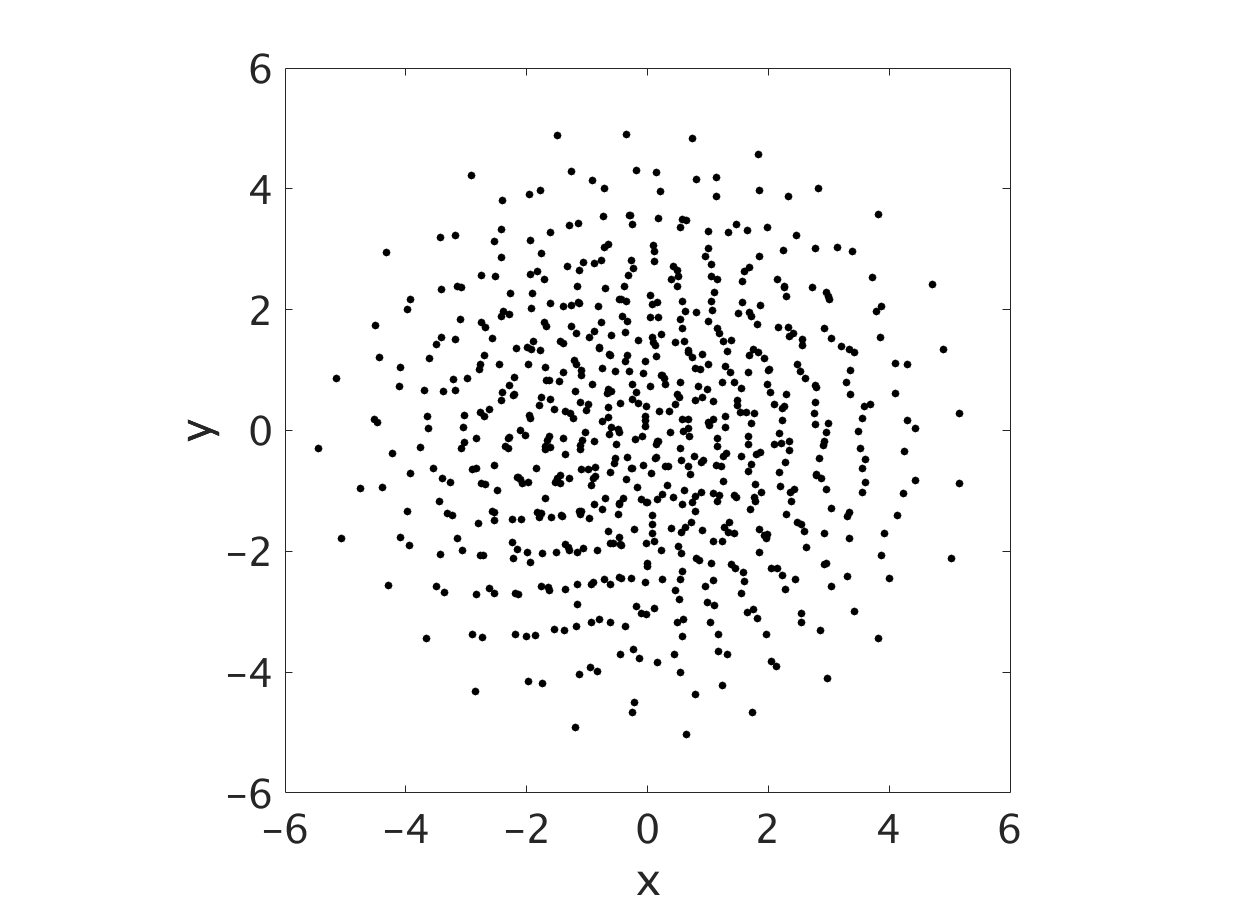}}
\caption{\label{fig:12and14e} The $xy$ projections of the two
  configurations in Fig.~\ref{fig:12and14dd} with 700 particles.  In
  the left panel, we have $L=6000$ and $A_{12}= 0.0024$.  In the right
  panel, we have $L=L_{\textrm{max}}= 7895$ and $A_{12}= 0.021$, here
  we see that the outer particles in the equator region are very
  loosely bound.  The asymmetry in this projection is too small to be
  visible.}
\end{center}
\end{figure}

The Figures~\ref{fig:12and14d}, \ref{fig:12and14dd},
and~\ref{fig:12and14e} show configurations of 400 and 700 particles,
with one value of the angular momentum below and one value equal to
the maximum value.  These figures confirm the visible features that we
defined in the Introduction as parts of the Jeans effect.  They are:
\begin{center}
  \begin{itemize}
  \item There is always circular symmetry in the rotation plane.
  \item Instability at the maximal angular momentum is due to particle
    loss from the centrifugal force at the equator.
  \item The equator becomes a sharp edge at the maximal angular momentum.
  \end{itemize}
\end{center}

The potential $\phi$, defined in Eq.~(\ref{eq:phi}) and plotted with
level curves in the figures, is everywhere negative, it goes to
$-\infty$ as $x\to\pm\infty$, and to zero along the $z$-axis as
$z\to\pm\infty$.  Note the two saddle points (Lagrange points) in each
plot, most visible in the top left panels.  With the corresponding
level curves they delimit the Roche lobe, within which an external
particle would be bound.  As $L$ increases towards $L_{\textrm{max}}$
the cloud of particles becomes unstable by filling its Roche lobe,
developing the sharp edge shaped by the Roche lobe.  For
$L<L_{\textrm{max}}$ there is no sharp edge.  These features are the
same for 400 and 700 particles.

We claim that, within their limits, these figures are good
representations of the behaviour of a polytropic gas.  As shown in
Appendix~\ref{app:laneemden}, the level curves of the pressure and density
inside a gravitationally bound polytrope are the same as the
equipotential surfaces (of gravitational plus centrifugal potential).
It should be remembered, however, that the correspondence between the
two different systems has its limits.  In particular, the potential
plotted in our figures has point sources and is not smooth like the
potential from the polytrope, and it includes our artificially
introduced short range repulsion potential.

Figure~\ref{fig:12and14e} illustrates both the absence of the Jacobi
transition, and the instability due to centrifugal forces, as
predicted by Jeans.  It shows the projections on the rotation plane or
$xy$-plane of the two configurations of 700 particles shown in
Fig.~\ref{fig:12and14dd}.  The configuration in the top panels is at
$L=6000$, and has rotational symmetry with no signs of the shedding of
particles. The configuration in the bottom panels is at
$L=L_{\textrm{max}}=7895$, and we see that the outermost particles
clearly are on the brink of drifting away.  But despite the fact that
the configuration is close to breakup, it remains rather circularly
symmetric, thus confirming the Jeans effect.

\section{Conclusion and outlook}

We have used a model with $N$ particles, where $N$ is a large but
finite number, to simulate a gravitationally bound rotating polytrope.
We argue that this method is valid, and that the continuum limit can
be described as the limit $N^{-1/3}\to 0$.

The method is useful because it is simpler than solving a partial
differential equation.  Although the continuum limit is approached
rather slowly, conclusions can be drawn from values of $N$ of a few
hundred, which are tractable numbers.

The polytropic index $n$ corresponds in our model to a repulsive
potential between the particles inversely proportional to the distance
to the power $3/n$.  We have studied only the special case $n=3/2$,
corresponding to an inverse square repulsive potential.

One of the problems left for further investigations is to change the
value of $n$.  A technical problem which then appears is that the
virial theorem becomes slightly more complicated, because the three
terms in Eq.~(\ref{eq:virthproof}) scale with three different powers
of $\alpha$.

In particular, it would be interesting to study the Jacobi transition
to a shape which is no longer rotationally symmetric in the rotation
plane, and to verify the necessary condition found by Jeans, that
$n<0.83$.

In a rather different direction, the method described here could be
used for studying rotating molecules, with totally different
interaction potentials.  The method used for computing stability can
be used for computing vibrational frequencies.

\appendix


\section{The Jeans theory of rotationally distorted polytropes}
\label{app:Jeans}

To find out how fast a body that is gravitationally compressed at its
centre can rotate before it becomes unstable, Jeans constructed what
he called the adiabatic model, described by an equation of state of
the form
\be
\label{eq:Jeanseqstate}
P=K\rho^{\gamma}-p_0\;.
\ee
We will summarize here very briefly his reasoning and main results.
See~\cite{JeansI,JeansII,JeansIII} for more details.

A positive constant $p_0$ in the equation of state implies that the
density $\rho$ at the surface, where $P=0$, may have a positive value
$\sigma$.  Physically, it means that the theory may describe an inner
part of a larger body.  He introduces a compressibiliy parameter
\be
\label{eq:epsdef}
\epsilon=\frac{\rho_c-\sigma}{\rho_c}\;,
\ee
where $\rho_c$ is the central density.  Thus, $\epsilon=0$ corresponds
to an incompressible fluid, or the central part of a larger body,
whereas $\epsilon=1$ for a polytrope with equation of state
$P=K\rho^{\gamma}$.

A slowly rotating incompressible body takes the shape of a Maclaurin
ellipsoid, or spheroid, with two equal axes.  With faster rotation
this becomes unstable, and bifurcates to a Jacobi ellipsoid, with
three different axes.  The shape of a slowly rotating compressible
body will be what Jeans calls a distorted spheroid, or
pseudo-spheroid.  A main result derived by Jeans is that the
bifurcation to a distorted Jacobi ellipsoid will not take place if the
adiabatic index $\gamma$ is too small, so that the body is too much
centrally condensed.  Then it becomes unstable instead by shedding
particles at the equator.

Jeans writes the variable density as
\be
\rho(x,y,z)=\rho_c\,(1-\epsilon F(x,y,z))\;.
\ee
The surface is given by the equation $\rho=\sigma$, and with the
definition~(\ref{eq:epsdef}) this means that $F=1$.  Knowing that the
surface is a spheroid in the incompressible case $\epsilon=0$, he
then writes
\be
F=\frac{x^2+y^2}{a^2}+\frac{z^2}{c^2}
+\epsilon P_0+\epsilon^2 Q_0+\epsilon^3 R_0+\cdots\;.
\ee
Here $a$ and $c$ are the semiaxes of the spheroid when $\epsilon=0$,
and $P_0,Q_0,R_0,\ldots$ are functions of $x,y,z$ describing how the
spheroid is distorted when $\epsilon>0$.  Expanding to second order in
$\epsilon$ he derives expressions for $P_0$ and $Q_0$ that take the
following form when $y=z=0$,
\be
P_0(x)=\frac{Lx^4}{a^8}+\frac{2px^2}{a^4}\;,\qquad
Q_0(x)=\frac{Rx^6}{a^{12}}+\frac{rx^4}{a^8}+\frac{2ux^2}{a^4}\;.
\ee
Here $L,p,R,r,u$ are coefficients that he determines by solving the
equations of hydrostatic equilibrium and gravitation.

On the $x$-axis, with $y=z=0$, to second order in $\epsilon$ the
surface is at $F(x)=1$, with
\be
F(x)=\frac{x^2}{a^2}+\epsilon P_0(x)+\epsilon^2 Q_0(x)\;.
\ee
The equation $F(x)=1$ may be written as
\be
\frac{x^2}{a^2}=1-\epsilon P_0(x)-\epsilon^2 Q_0(x)\;,
\ee
and solved by iteration.  To second order in $\epsilon$ this gives
that
\be
\label{eq:x2a2}
\frac{x^2}{a^2}=1
-\epsilon{\left[\frac{L}{a^4}+\frac{2p}{a^2}\right]}
+\epsilon^2{\left[
     \frac{2L^2}{a^8}
    +\frac{6Lp-R}{a^6}
    +\frac{4p^2-r}{a^4}
    -\frac{2u}{a^2}\right]}\;.
\ee

The critical condition that the centrifugal force is equal to the
gravitational force at the equator, is expressed by the condition
that the derivative of the pressure vanishes,
\be
\dd{P(x)}{x}=0\;,
\ee
at the point where $F(x)=1$.  The equation of state $P=K\rho^{\gamma}$
means that the equations $P'(x)=0$, $\rho'(x)=0$, and $F'(x)=0$ are
equivalent.  We have that
\be
\frac{a^2}{2x}\,F'(x)=
1+\epsilon{\left[\frac{2Lx^2}{a^6}+\frac{2p}{a^2}\right]}
+\epsilon^2{\left[\frac{3Rx^4}{a^{10}}+\frac{2rx^2}{a^6}
    +\frac{2u}{a^2}\right]}\;.
\ee
Inserting $x$ from Eq.~(\ref{eq:x2a2}) we get, to second order in
$\epsilon$,
\be
\frac{a^2}{2x}\,F'(x)=
1+\epsilon{\left[\frac{2L}{a^4}+\frac{2p}{a^2}\right]}
+\epsilon^2{\left[
    -\frac{2L^2}{a^8}
    +\frac{3R-4Lp}{a^6}
    +\frac{2r}{a^4}
    +\frac{2u}{a^2}\right]}\;.
\ee

The two equations $F(x)=1$ and $F'(x)=0$ together determine the point
where the rotation becomes so fast that the body starts losing
particles at the equator.  Now Jeans wants to compare this to the
point where the transition from a pseudo-spheroidal to a
pseudo-elliptical shape takes place.  Through a lengthy analysis he
finds numerical values for the coefficients $L,p,R,r,u$, depending on
the adiabatic index $\gamma$, at the transition point~\cite{JeansIII}.
With these values, to second order in $\epsilon$ the equation
$F'(x)=0$ at the surface where $F(x)=1$ takes the form
\be
1+\epsilon{\left[(\gamma-2)-1.0509\right]}
+\epsilon^2{\left[
    \frac{1}{2}(\gamma-2)^2-0.4063(\gamma-2)-0.0510\right]}=0\;.
\ee
Setting $\epsilon=1$, the value for a polytrope with density $\rho=0$
at the surface, we get to first order in $\epsilon$ the critical value
$\gamma=2.0509$, and to second order $\gamma=2.1521$.  Hence Jeans
guesses that the values for higher order approximations may converge
to a limit
\be
\gamma_c=1+\frac{1}{n_c}\approx 2.2\;,
\ee
which means that the critical polytropic index will be
$n_c\approx 0.83$.

In conclusion, the transition where the shape changes may take place
only if the polytropic index $n$ is smaller than $n_c\approx 0.83$.
For larger values of $n$, meaning higher compressibility, the
mechanism for instability of the pseudo-spheroid will be shedding of
particles from the equator.

Jeans also computed the shape of the rotating body at the critical
speed of rotation where the particle loss sets in.  The shape is then
a pseudo-spheroid with a sharp edge at the equator, as sketched
in~\cite{JeansII}, Fig.~43.  In our simulations the same shape is
apparent in Fig.~\ref{fig:12and14d}.

\section{The Lane--Emden equation}
\label{app:laneemden}

The equation of hydrostatic equilibrium in a rotating reference system
is
\be
\label{eq:hydrostat}
\nabla P=-\rho\,\nabla(\phi_g+\phi_c)\;,
\ee
where $\phi_g$ is the gravitational potential and $\phi_c$ the centrifugal
potential,
\be
\label{eq:VgVc}
\nabla^2 \phi_g=4\pi G\rho\;,\qquad
\phi_c=-\frac{1}{2}\,\Omega^2(x^2+y^2)\;.
\ee
Here $G$ is the gravitational constant and $\Omega$ is the angular
velocity.  We take the rotation axis as our $z$ axis.  There is the
boundary condition $\phi_g=0$ at infinity.  Eq.~(\ref{eq:polystate})
implies that
\be
\frac{\nabla P}{\rho}
=\nabla\left(K(n+1)\,\rho^{1/n}\right).
\ee
Hence inside the body, where $\rho>0$, Eq.~(\ref{eq:hydrostat}) may
be integrated to give that
\be
K(n+1)\,\rho^{1/n}=-\phi_g-\phi_c+\phi_0\;.
\ee
The potentials $\phi_g$ and $\phi_c$ are given by Eq.~(\ref{eq:VgVc}), and
$\phi_0$ is an integration constant such that $\phi_0=\phi_g+\phi_c$ where
$\rho=0$, on the surface of the body.  We now write
\be
\rho=\rho_c\,\theta^n\;,
\ee
where $\rho_c$ is the central density and $\theta$ is dimensionless,
$\theta=1$ at the centre and $\theta=0$ at the surface.  Inside the
body, where $\theta>0$, we have that
\be
\label{eq:basic}
K(n+1)\,\rho_c^{1/n}\,\theta=-\phi_g-\phi_c+\phi_0\;.
\ee
By differentiating this equation we get that
\be
K(n+1)\rho_c^{1/n}\,\nabla^2\theta
=-4\pi G\rho_c\,\theta^n+2\Omega^2\;.
\ee
The differentiation introduces lots of unphysical solutions.  The
physically meaningful solutions are those that are also solutions of
Eq.~(\ref{eq:basic}).

By a suitable scaling of the coordinates $x,y,z$ we arrive at the
dimensionless equation
\be
\label{eq:ChaMilne}
\nabla^2\theta=-\theta^n+2\omega^2\;,
\ee
where $\omega$ is a scaled angular velocity.  The special case
$\omega=0$ is the Lane--Emden equation.

The physically meaningful solutions of the Lane--Emden equation are
those that are spherically symmetric.  Therefore we take
$\theta=\theta(\xi)$, where $\xi$ is a dimensionless radius, and
arrive at the following standard form of the equation,
\be
\label{eq:LaneEmden}
\frac{1}{\xi^2}\,\dd{}{\xi}\left(\xi^2\,\dd{\theta}{\xi}\right)
=-\theta^n\;.
\ee

With initial conditions $\theta(0)=1$ and $\theta'(0)=0$ it describes
the density profile of a gravitationally bound nonrotating polytropic
gas of given polytropic index $n$.  The initial conditions make
$\theta$ an even function of $\xi$.  The density at a radius $r=a\xi$
is
\be
\rho(r)=\rho_c\,\theta^n(\xi)\;,
\ee
where $\rho_c$ is the central density and $a$ is a scaling factor.
When $n<5$ there is a sharp surface at some value $\xi=\xi_1$ where
$\theta(\xi_1)=0$.

\subsubsection*{The truncated Taylor series}

We want the solution for $n=3/2$.  It is known that the Taylor
expansion
\be
\theta(\xi)=\sum_{k=1}^K a_k\,\xi^{2(k-1)}
\ee
with $K=\infty$ converges all the way to $\xi=\xi_1$
\cite{RoxburghStockman}.  Taking $K$ to be finite we obtain an
approximate solution which is a polynomial of degree $2(K-1)$.  The
coefficients $a_k$ are rational numbers, easily determined by some
computer algebra program.  The first 17 coefficients, starting with
$a_1=1$, are as follows.
\be
a_2=-\frac{1}{6}\;,\quad
a_3= \frac{1}{80}\;,\quad
a_4=-\frac{1}{1440}\;,\quad
a_5= \frac{1}{31104}\;,\quad
a_6=-\frac{19}{14256000}\;,\quad
\ee

\be
\begin{array}{rrr}
a_7 =  5.09419635577042984\,e-8\;,\phantom{0}&&
a_8 = -1.83974190532832508\,e-9\;,\phantom{0}\\
a_9 =  6.34122309412795978\,e-11\;,&&
a_{10} = -2.11355232821624019\,e-12\;,\\
a_{11} =  6.82513007546544624\,e-14\;,&&
a_{12} = -2.16016381861918138\,e-15\;,\\
a_{13} =  6.66006333781310159\,e-17\;,&&
a_{14} = -2.03829508290751622\,e-18\;,\\
a_{15} =  6.04226394127875035\,e-20\;,&&
a_{16} = -1.82280044869048828\,e-21\;,\\
a_{17} =  5.13479258506338056\,e-23\;.
\end{array}
\ee
The first root of the polynomial of degree 32 is at
\be
\xi_1=3.653\,853\,288\,284\,8\;.
\ee

We introduce here a scaled function 
\be
f(u)=\alpha\,\theta(\beta u)=\sum_{k=1}^K b_k\,u^{2(k-1)}\;.
\ee
With $\alpha=\xi_1^{\,4}$ and $\beta=\xi_1$ this is a solution
of the Lane--Emden equation
\be
\label{eq:LaneEmdenII}
\frac{1}{u^2}\,\dd{}{u}\left(u^2\,\dd{f}{u}\right)
=-f^{3/2}
\ee
on the interval $0\leq u\leq 1$, with $f(1)=0$.

With the even Taylor series the left hand side of
Eq.~(\ref{eq:LaneEmdenII}) is nonsingular, it is
\be
L(u)=\sum_{k=2}^K 2(k-1)(2k-1)\,b_k\,u^{2(k-2)}\;.
\ee
With $K$ finite the residual of the equation, the left hand minus the
right hand side, is
\be
\Delta(u)=L(u)+(f(u))^{3/2}\;.
\ee
The logarithm of this is plotted for $K=17$ in
Fig.~\ref{fig:laneemdenI}.  It is zero to numerical precision for
$u<0.4$, then it turns positive and grows nearly exponentially to
$\Delta(1)=5.48$.  The root mean square residual is $0.615$.

\subsubsection*{Better polynomial approximations}

\begin{figure}
\begin{center}
\fbox {\includegraphics[width=7.48cm]{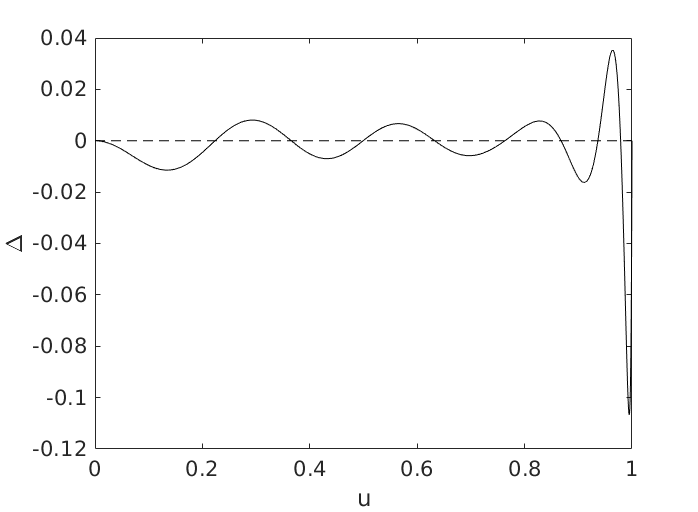}}
\fbox {\includegraphics[width=7.48cm]{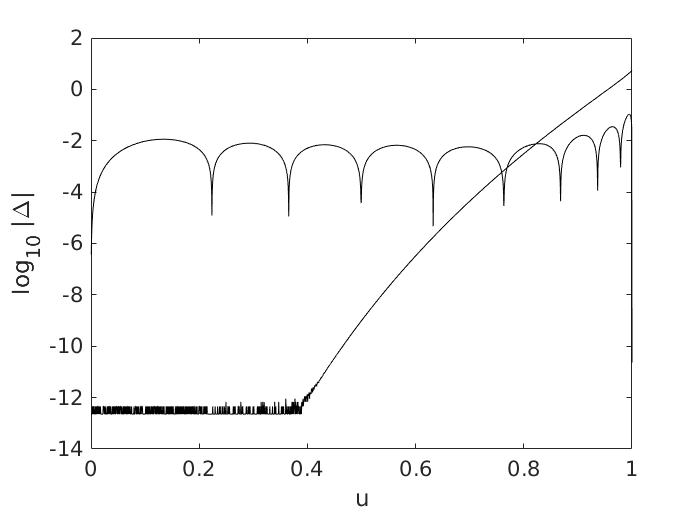}}
\caption{\label{fig:laneemdenI} The left panel shows the residual
  $\Delta(u)$ of the Lane--Emden equation with our polynomial
  approximation to the solution.  The right panel shows the same
  residuals on a logarithmic scale, compared with the residuals
  obtained with our truncated Taylor series.}
\end{center}
\end{figure}

The truncated Taylor series is not necessarily the best possible
polynomial approximation.  We present here a polynomial of degree 24,
\be
f(u)=\sum_{k=1}^{13} c_k\,u^{2(k-1)}\;,
\ee
which gives smaller residuals for $u>0.8$, as shown in
Fig.~\ref{fig:laneemdenI}.  We see that $\Delta(u)$ has eight zeros
for $0<u<1$, and $-0.107<\Delta(u)<0.036$ for all $u$.

We impose three constraints on the coefficients $c_k$.  We require
that
\be
\label{eq:f1is0}
f(1)=\sum_{k=1}^{13} c_k=0\;.
\ee
Then we require that the residual $\Delta(u)$ vanishes at $u=0$ and
$u=1$.  The equation $\Delta(0)=0$ holds when
\be
\label{eq:c2fromc1}
c_2 = -\frac{c_1^{\,3/2}}{6}\;.
\ee
For $f(1)=0$ the equation $\Delta(1)=0$ holds when
\be
L(1)=\sum_{k=2}^{13} 2(k-1)(2k-1)\,c_k=0\;.
\ee

We use the following values for the coefficients, found by an
approximate minimization of the sum of the residuals squared.  We do
not claim that they are optimal values.  The root mean square residual
is $0.013$.
\be
\begin{array}{rrr}
c_1    = \phantom{-}178.220\,339\,615\;,&&
c_2    =           -396.537\,852\,238\;,\\
c_3    = \phantom{-}396.955\,994\,705\;,&&
c_4    =           -292.916\,881\,695\;,\\
c_5    = \phantom{-}168.056\,325\,560\;,&&
c_6    =        \;\,-29.041\,745\,736\;,\\
c_7    =           -181.962\,696\,622\;,&&
c_8    = \phantom{-}480.394\,543\,252\;,\\
c_9    =           -728.865\,230\,225\;,&&
c_{10} = \phantom{-}723.088\,520\,738\;,\\
c_{11} =           -459.489\,817\,966\;,&&
c_{12} = \phantom{-}169.976\,902\,778\;,\\
c_{13} =        \;\,-27.878\,402\,166\;.
\end{array}
\ee
Of these 13 coefficients, only 10 are independent.  For example, given
$c_1$ and $c_5,c_6,\ldots,c_{13}$ we compute $c_2$ by
Eq.~(\ref{eq:c2fromc1}).  Then we compute
\be
s_1&\!\!\!=&\!\!\!c_1+c_2+\sum_{k=5}^{13}c_k\;,\nonumber\\
s_2&\!\!\!=&\!\!\!   6c_2+\sum_{k=5}^{13}2(k-1)(2k-1)c_k\;,
\ee
and
\be
c_3&\!\!\!=&\!\!\!\frac{s_2-42s_1}{22}\;,\nonumber\\
c_4&\!\!\!=&\!\!\!\frac{20s_1-s_2}{22}\;.
\ee
The value of $c_1$ given here corresponds to a zero of the Lane--Emden
function $\theta(\xi)$ which is
\be
\xi_1=c_1^{\,1/4}=3.653\,754\,109\;.
\ee
The residuals of the equation are plotted in
Fig.~\ref{fig:laneemdenI}.

\subsubsection*{The mass $M(r)$ within radius $r$}

\begin{figure}
\begin{center}
\fbox {\includegraphics[width=8cm]{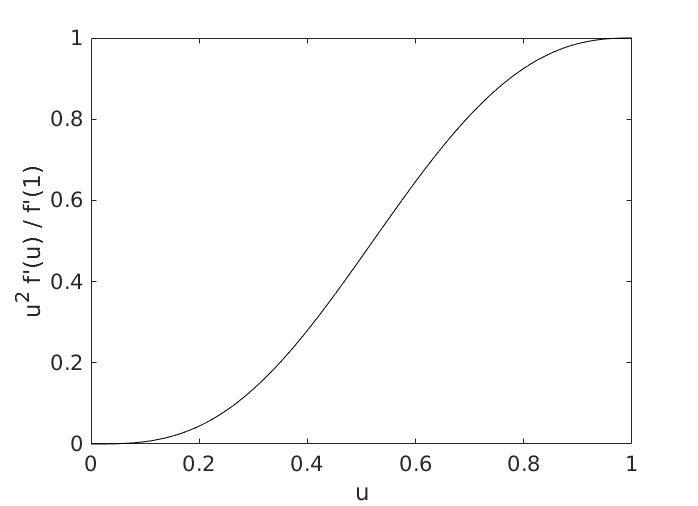}}
\caption{\label{fig:Mr} The mass fraction as a function of $u$.}
\end{center}
\end{figure}

If $\rho(r)=\lambda\,(f(u))^{3/2}$ is the mass density at a radius
$r=bu$, with $\lambda=\rho_c/\alpha^{3/2}$, then the total mass within
$r$ is
\be
M(r)
=4\pi\lambda b^3\int_0^u \rmd v\;v^2\,(f(v))^{3/2}
=-4\pi\lambda b^3\,u^2f'(u)\;.
\ee
We use Eq.~(\ref{eq:LaneEmdenII}).  Here $f'(u)<0$ for $u>0$.  The
fraction of the total mass within $r=bu$ is
\be
\label{eq:Fu}
F(u)=\frac{u^2f'(u)}{f'(1)}\;.
\ee
This is plotted in Fig.~\ref{fig:Mr}.

The method for generating one random point from the Lane--Emden
density distribution is to generate a random number $w\in{[0,1]}$ and
solve the equation $F(u)=w$ for $u$.  The radius is $r=bu$, with a
scaling factor $b$ to be determined later.  The $x,y,z$ coordinates
are
\be
x=r\sin\vartheta\,\cos\varphi\;,\qquad
y=r\sin\vartheta\,\cos\varphi\;,\qquad
z=r\cos\vartheta\;.
\ee
We generate uniform random variables $w_1,w_2\in{[0,1]}$ and define
\be
\cos\vartheta=2w_1-1\;,\qquad
\sin\vartheta=\sqrt{1-\cos^2\vartheta}\;,\qquad
\varphi=2\pi w_2\;.
\ee

We used this Monte Carlo method to generate 10\,000 points.  For every
pair of points we compute their mutual distance $d$.  We compute the
scaling factor $b$ by minimizing the energy, Eq.~(\ref{eq:Esumpairs}),
or equivalently by requiring the virial theorem to hold,
Eq.~(\ref{eq:virth1}).  Repeating the calculation several times, we
get a mean value with a statistical error,
\be
b=5.2760\pm 0.0080\;.
\ee
Since by definition $u\leq 1$, $b$ is the outer radius of the cloud of
10\,000 particles.  The same Monte Carlo data give a root mean square
radius which is
\be
r_m=2.9220\pm 0.0029\;,
\ee
and an average distance between points which is
\be
\langle d\rangle=3.8432\pm 0.0037\;.
\ee
The average energy per particle pair is
\be
\langle E_2\rangle=-0.16253\pm 0.00015\;.
\ee

\subsubsection*{Note on numerical methods}

We write $f(u)=g(v)$ with $v=u^2$.  The derivatives of these functions
are
\be
f'(u)  = 2ug'(v)\;,\qquad
f''(u) = 4vg''(v)+2g'(v)\;.
\ee
The following Matlab function computes $f,f',f''$ and (internally
only) $g,g',g''$.  It also computes the residual $\Delta$ of the
Lane--Emden equation.  The input argument $u$ may be a matrix of any
size, then the outputs $f,f',f'',\Delta$ are matrices of the same
size.  Before calling the function we need to declare the coefficients
cf to be global and assign values to them.  Note that ``$.*$''
denotes elementwise multiplication of matrices.

\begin{figure}[h]
  Matlab code for computing the function $f(u)$, its derivatives, and
  the residue of Eq.~(\protect{\ref{eq:LaneEmdenII}}).
\begin{lstlisting}
  function [f,f1,f2,Delta] = ff(u)
  global cf
  v  = u.*u;
  g  = cf(12)+cf(13)*v;
  g1 = cf(13);
  g2 = 0;
  for k=2:12
    g2 = 2*g1+g2.*v;
    g1 = g+g1.*v;
    g  = cf(13-k)+g.*v;
  end
  g  = max(g,0);
  f  = g;
  f1 = 2*u.*g1;
  f2 = 4*v.*g2+2*g1;
  Delta = f2+4*g1+g.^(3/2);
\end{lstlisting}
\end{figure}


\begin{figure}[h]
  Matlab code for computing the mass fraction $F(u)$,
  and the inverse of this function.
\begin{lstlisting}
  function OUT = F(u)
  [f1,fd1] = ff(1);
  [f,fd] = ff(u);
  OUT = u.*u.*fd/fd1;

  function OUT = Finv(v)
  u1 = 0;
  u2 = 1;
  u  = 10;
  ua = 20;
  while (u%*$\sim$*)=ua)
    ua = u;
    u = 0.5*(u1+u2);
    if (F(u)<v)
      u1 = u;
    else
      u2 = u;
    end
  end
  OUT = u;
\end{lstlisting}
\end{figure}

\pagebreak

\bibliographystyle{unsrt}

\begin{thebibliography}{99}
 
\bibitem{HM}
Y.~Hopstad and J.~Myrheim,\\
{\em Computer simulations of rotating systems of few particles bound
  by gravitation.}\\
Int.~J.~Mod.~Phys.~C 27, 1650142 (2016).

\bibitem{ChandraStellarStr} 
S.~Chandrasekhar,
{\em An Introduction to the Study of Stellar Structure.}\\
Dover edition (1958).

  
\bibitem{Emd} 
R.~Emden,
{\em Gaskuglen.}
Verlag B.G.~Teubner, Leipzig und Berlin (1907).   
 
\bibitem{Edd} 
A.S.~Eddington,
{\em The Internal Constitution of the Stars.}\\
Cambridge University Press (1926). 
 
\bibitem{Mil} 
E.A.~Milne,
{\em The Equilibrium of a Rotating Star.}\\
Monthly Notices of the Royal Astronomical Society 83, 118 (1923). 

\bibitem{Esp}
F.L.~Espinosa and M.~Rieutord,
{\em Gravity Darkening in Rotating Stars.}\\
Astron.~Astrophys.~533, A43 (2011).


\bibitem{Cha} 
S.~Chandrasekhar,
{\em The Equilibrium of Distorted Polytropes.}\\
Monthly Notices of the Royal Astronomical Society
93, 390-405 (1933). 

  

\bibitem{JeansI} 
J.H.~Jeans,
{\em Problems of Cosmogony and Stellar Dynamics.}\\
Cambridge University Press (1919).

\bibitem{JeansII} 
J.H.~Jeans,
{\em Astronomy and Cosmogony.}\\
Cambridge University Press (1929).

\bibitem{JeansIII}
  J.H.~Jeans,\\
  {\em Bakerian Lecture 1917: The Configurations of Rotating
    Compressible Masses.}\\
  Philosophical Transactions of the Royal Society of London, Series
  A.~218, 157-210 (1919).

\bibitem{Kho}
K.V.~Kholshevnikov,
{\em On the Lyapunov Theory of Equilibrium Figures
  of Celestial Bodies.}\\
Vestnik St.~Petersburg University Mathematics 40, 123 (2007).

\bibitem{James}
R.A.~James,
{\em The Structure and Stability of Rotating Gas Masses.}\\
Astrophys.~J.~140, 552 (1964).

\bibitem{Rox}
J.J.~Monaghan and I.W.~Roxburgh,
{\em The Structure of Rapidly Rotating Polytropes.}\\ 
Monthly Notices of the Royal Astronomical Society 131, 13-22 (1965). 

\bibitem{RobIandII}
P.H.~Roberts,
{\em On Highly Rotating Polytropes. I, II}\\
Astrophys.~J.~137, 1129 and~138, 809 (1963).

\bibitem{RobIII}
M.~Hurley and P.H.~Roberts,
{\em On Highly Rotating Polytropes. III}\\
Astrophys.~J.~140, 583 (1964).

\bibitem{Kongetal}
D.~Kong, K.~Zhang, and G.~Schubert,\\
{\em Self-consistent internal structure of a rotating gaseous planet
  and its comparison with an approximation by oblate spheroidal
  equidensity surfaces.}\\
Physics of the Earth and Planetary Interiors 249, 43 (2015).




\bibitem{MRFELBP}
M.~Rieutord, F.~Espinosa Lara, and B.~Putigny,\\
{\em An algorithm for computing the 2D structure of fast rotating stars.}\\
J.~Comput.~Phys.~318, 277 (2016).


\bibitem{DGREG}
  D.~Gondek-Rosinska and E.~Gourgolhon,\\
  {\em Jacobi-like bar mode instability of relativistic rotating bodies.}\\
Phys.~Rev.~D 66, 044021 (2002).


\bibitem{CR}
P.H.~Chavanis and M.~Rieutord,\\
{\em Statistical mechanics and phase diagrams of rotating
  self-gravitating fermions.}\\
Astronomy and Astrophysics 412, 1 (2003).



\bibitem{Shap}
S.L.~Shapiro and S.A.~Teukolsky,\\
{\em Black Holes, White Dwarfs, and Neutron Stars.}
Wiley (1983).


\bibitem{Tas} 
J.-L.~Tassoul,
{\em Theory of Rotating Stars.}
Princeton University Press (1978).

\bibitem{RoxburghStockman}
I.W.~Roxburgh and L.M.~Stockman,
{\em Power series solutions of the polytrope equations.}\\
Monthly Notices of the Royal Astronomical Society 303, 466 (1999).

\end{thebibliography}

\end{document}